\documentclass[11pt,fleqn]{article}  % OK for EJC

\usepackage{amsmath,amssymb}
\usepackage{graphicx}
\usepackage{ignore}
\usepackage{subfigure}

\newif\ifTA   % Think Aloud
  \TAfalse
\newcommand{\du}{\Delta u}
\newcommand{\dtilu}{\Delta\tilde{u}}

\newtheorem{remark}{Remark}
\newtheorem{TA}{Thinking Aloud}

\newtheorem{algorithm}{Algorithm}
\newtheorem{theorem}{Theorem}
\newtheorem{lemma}{Lemma}

\begin{document}
%\begin{frontmatter}


\title{Multiplexed Model Predictive Control
\thanks{This work is supported by A*STAR project ``Model Predictive
Control on a Chip'' (Ref: 052-118-0059). Much of the material in
this paper was previously presented at the IFAC World Congress 2005,
and at the European Control Conference 2007. KV Ling would also like
to acknowledge the support of Tan Chin Tuan Fellowship and of
Pembroke College, Cambridge.}}
\author{Keck Voon Ling\\ %(corresponding author)\\
    School of Electrical and Electronics Engineering\\
      Nanyang Technological University\\
      Singapore, 639798\\ \texttt{ekvling@ntu.edu.sg}
      \and
Jan Maciejowski\\Department of Engineering\\ University of
    Cambridge\\ United Kingdom
    \and
Arthur Richards\\Department of Aerospace Engineering\\ University of
    Bristol\\ United Kingdom
    \and
Bing Fang Wu\\
    School of Electrical and Electronics Engineering\\
      Nanyang Technological University\\
      Singapore
 }

%\date{14 October 2008}
\date{16 February 2010}

% Cover page for Technical Report:
\begin{center}

\rule{0mm}{7cm} % strut
MULTIPLEXED MODEL PREDICTIVE CONTROL

\bigskip
K.V. Ling, J.M. Maciejowski, A.G. Richards, and B-F. Wu

\bigskip
UNIVERSITY OF CAMBRIDGE

\bigskip
Department of Engineering

\bigskip
TECHNICAL REPORT CUED/F-INFENG/TR.657

16 February 2010
\thispagestyle{empty} % no number on cover page
\end{center}
% End of cover page

\maketitle
\pagenumbering{arabic} % reset page counter to 1

\begin{abstract}
This paper proposes a form of MPC in which the control variables are
moved asynchronously. This contrasts with most MIMO control schemes,
which assume that all variables are updated simultaneously.
%Most academic control schemes for MIMO systems assume all the
%control variables are updated simultaneously.
MPC outperforms other control strategies through its ability to deal
with constraints. This requires on-line optimization, hence
computational complexity can become an issue when applying MPC to
complex systems with fast response times. The multiplexed MPC scheme
described in this paper solves the MPC problem for each subsystem
sequentially, and updates subsystem controls as soon as the solution
is available, thus distributing the control moves over a complete
update cycle. The resulting computational speed-up allows faster
response to disturbances, which may result in improved performance,
despite finding sub-optimal solutions to the original problem.
%The multiplexed MPC scheme is also closer to
%industrial practice in many cases. This paper presents initial
%stability results for multiplexed MPC.
%%KVL010706
%An interesting connection between the Multiplexed MPC and the Gauss-Seidel method
%of iteratively solving a system of linear equation is also
%established. Some preliminary work on comparing Multiplexed MPC
%with conventional MPC are also reported.
\end{abstract}

%\begin{keyword}

%\end{frontmatter}

% Bibliographic references with the natbib package:
% Parenthetical: \citep{Bai92} produces (Bailyn 1992).
% Textual: \citet{Bai95} produces Bailyn et al. (1995).
% An affix and part of a reference:
%   \citep[e.g.][Ch. 2]{Bar76}
%   produces (e.g. Barnes et al. 1976, Ch. 2).

\section{Introduction} \label{intro}

\subsection{The basic idea} \label{basicidea}

Model Predictive Control (MPC) has become an established control
technology in the petrochemical industry, and its use is currently
being pioneered in an increasingly wide range of process industries
\cite{QB03,VirWanRob04}. It is also being proposed for a range of
higher bandwidth applications, such as ships \cite{PeGoTz:00},
aerospace \cite{MuHaJa:03,RiHo:03}, and road vehicles
\cite{MoBaBo:03}. This paper is concerned with facilitating
applications of MPC in which computational complexity, in particular
computation time, is likely to be an issue. One can foresee that
applications to embedded systems, with the MPC algorithm implemented
in a chip or an FPGA
\cite{Bleris:06,Johanson:06,KnaWilMilNin09,LingYueMac:06}, are
likely to run up against this problem.
%have this characteristic.

\begin{figure}[htb]
  \begin{center}
  \includegraphics[width=0.8\textwidth]{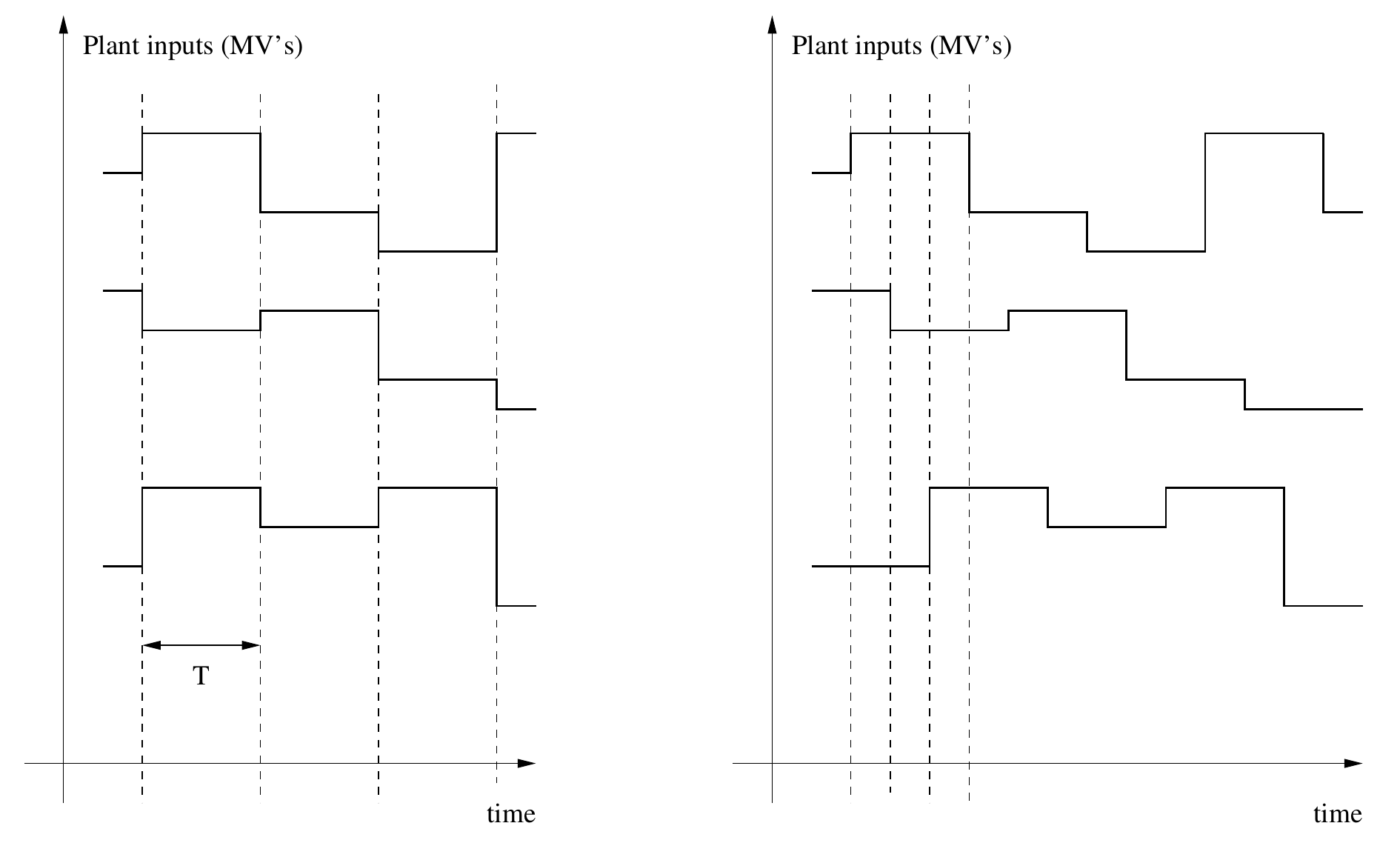}
  \caption{Patterns of input moves for conventional `synchronized'
  MPC (left), and for the Multiplexed MPC (right) introduced in this paper.}
  \label{fig:inputpattern}
  \end{center}
\end{figure}

MPC operates by solving an optimization problem on-line, in real
time,  to determine a plan for future operation. Only an initial
portion of that plan is implemented, and the process is repeated,
re-planning when new information becomes available. Since numerical
optimization naturally handles hard constraints, MPC~offers good
performance while operating close to constraint
boundaries~\cite{jmmbook}. Solving a numerical optimization can be a
complex problem, and for situations in which computation is limited,
the time to find the solution can be the limiting factor in the
choice of the update interval. Most MPC theory to date, and as far
as we know all implementations, assumes that all the control inputs
are updated at the same instant. Suppose that a given MPC control
problem can be solved in not less than $T$ seconds, so that the
smallest possible update interval is $T$. The computational
complexity of typical MPC problems, including time requirements,
tends to vary as $O((m\times N_u)^3)$, where $m$ is the number of
control inputs and $N_u$ is the horizon length. We propose to use
MPC to update only one control variable at a time, but to exploit
the reduced complexity to update successive inputs at intervals
smaller than $T$, typically $T/m$. After $m$ updates a fresh cycle
of updates begins, so that each whole cycle of updates repeats with
cycle time $T$. We call this scheme \emph{multiplexed MPC}, or MMPC.
We assume that fresh measurements of the plant state are available
at these reduced update intervals $T/m$. The main motivation for
this scheme is the belief that in many cases the approximation
involved in updating only one input at a time will be outweighed
--- as regards performance benefits --- by the more rapid response
to disturbances, which this scheme makes possible. It is often the
case that ``do something sooner'' leads to better control than ``do
the optimal thing later''. %In this paper we focus on establishing
%closed-loop stability of particular variations of multiplexed MPC.
%In a later paper we shall consider the performance benefits.
Fig. \ref{fig:inputpattern} shows the pattern of input moves in the
MMPC scheme with $m=3$, compared with the conventional scheme in
which the three input moves are synchronized. We will refer to
conventional MPC as \emph{Synchronized MPC}, or SMPC, in the rest of
this paper.

%A specific example of an application in which our scheme could be effective is the roll
%stabilisation of ships. Typically rotatable fins are installed for the specific purpose of opposing
%the rolling torque that arises from wave actions. But the ship's rudder can also produce a rolling
%torque, and so is available as a second actuator for roll stabilization~\cite{PeGoTz:00}. This is
%useful for combatting the effects of unusually large waves, or for reducing the mean-square roll
%angle in heavy seas, when the fins may be saturated in rate or angle, and thus leads to an interest
%in designing a roll controller which uses both actuators. If the interval between actuator actions
%is $T$, then the ship's roll is essentially uncontrolled during intervals of duration $T$. Now if
%the roll angle is considered to be a band-limited stochastic process, such that its spectrum
%$S_{\phi\phi}(\omega)=0$ for $\omega>\Omega$, then the mean-square variation of the roll angle
%between control actions is limited by $T^2$~\cite{Pap:84}:
%\begin{equation}\label{BLroll}
%    E\{[\phi(t+T)-\phi(t)]^2\} \leq \Omega T^2 E\{\phi(t)^2\}
%\end{equation}
%Thus reducing $T$ offers the possibility of reducing the controlled mean-square roll angle. In
%order to reduce $T$ the possibility of controlling the fins and the rudder by means of our
%multiplexed MPC scheme could be considered.

The scheme which we investigate here is close to common industrial
practice in complex plants, where it is often impossible to update
all the control inputs simultaneously, because of their sheer
number, and the limitations of the communications channels between
the controller and the actuators.

In addition to treating the `nominal' MMPC case, in which the model
is assumed to represent the plant perfectly, we extend MMPC to
guarantee robust constraint satisfaction and feasibility of all
optimizations despite the action of unknown but bounded
disturbances. These are key issues in MPC: performance benefits are
achieved by operating close to constraint boundaries, but when the
state evolution no longer matches the predictions, constraint
violation and infeasibility can result. Many methods have been
developed to endow conventional synchronous MPC with
robustness~\cite{MayRawRaoSco:00,BM99a}. For use with MMPC, we have
adopted the \emph{constraint tightening}
approach~\cite{GKR97,CRZ01,RH06a,R05}, in which the constraints of
the optimization are modified to retain a margin for future feedback
action. Since only the constraint limits are modified, the
computational complexity remains the same as for the equivalent
nominal MPC. Constraint tightening is therefore well-suited to MMPC,
which is aimed at computation-limited applications.

Various generalizations of our scheme are possible. For example,
subsets of control inputs might be updated simultaneously, perhaps
all the inputs in each subset being associated with one subunit.
%Alternatively, sensor outputs might become available one at a time
%(or one subset at a time), as in the common practice of ``polling''
%sensors.
The assumption of equal intervals between the updates of plant
inputs is not essential to the MMPC idea. Any pattern of update
intervals can be supported, providing that it repeats in subsequent
update cycles. A further generalization, albeit involving a
significantly harder problem, would be not to update each control
input in a fixed sequence, but to decide in real time which input
(if any) needs updating most urgently --- one could call this
\emph{just-in-time MPC}.
%; note that this would then resemble \emph{statistical process
%control (SPC)}, which is used widely in manufacturing processes
%\cite{BoLu:97}.

\subsection{Related Work}
\label{related}

MMPC is related to distributed MPC~(DMPC)~\cite{CJKT02}, both
dividing the optimisation into smaller sub-problems. Several works
have been published which propose `distributed MPC' in the sense
that subsets of control inputs are updated by means of an MPC
algorithm. But these usually assume that several sets of such
computations are performed in parallel, on the basis of local
measurements only, and that all the control inputs are then updated
simultaneously. In some applications, such as formation flying of
unmanned vehicles \cite{DM06}, it is assumed that the state vectors
of subunits (vehicles) are distinct, and that coupling between
subunits occurs only through constraints and performance measures.
In \cite{VeRaWr:04} five different MPC-based schemes are proposed,
of which four are distributed or decentralized MPC schemes of some
kind. Their schemes 4 and 5 are the closest to our multiplexed
scheme. In these schemes an MPC solution is solved iteratively for
each control input, but it is assumed that no new sensor information
arrives during the iteration, and that all the control inputs are
updated simultaneously when the iterations have been completed.

Various approaches to robust DMPC have been investigated, including
worst-case predictions~\cite{JK02}, retention of ``emergency''
plans~\cite{SHF04,BKB04}, invariant ``tube'' predictions~\cite{TR06}
and constraint tightening~\cite{RH04acc}. Work on DMPC has typically
focussed on spatially distributed systems with some structure in the
system, \emph{e.g.}~teams of vehicles with decoupled dynamics. In
contrast, our new robust MMPC makes no assumptions on the overall
system structure, and considers temporal distribution, breaking the
optimisation down into a sequence of smaller problems, potentially
on the same processor.

%%%  START OF JMM ADDITIONS 13.04.07:
In \cite{BenCelHam06} a similar scheme to ours is proposed, but it
is assumed that a limitation occurs on network bandwidth rather than
on central computing resources. Hence optimal trajectories are
computed for all the plant inputs, but these are communicated to the
plant one input (or one group of inputs) at a time --- with the
optimization taking this communication restriction into account. If
the communication sequence is fixed and periodic then the scheme
proposed in \cite{BenCelHam06} is essentially the same as a version
of MMPC to which we previously referred as `scheme 1', except that
we allowed constraints on inputs and states \cite{LinMacWu05}.
\cite{BenCelHam06} also considers the case that a feedback law is
fixed for each input (or group of inputs), that the inputs are
updated according to some periodic scheme, and that a heuristic is
used to determine (online) the best point in the period for a given
state; this gives a heuristic version of `just-in-time control' as
defined above, though not really MPC any longer, since the feedback
law is assumed to be predetermined.
%We emphasise that, while
%\cite{BenCelHam06} and several of the references cited therein
%assume the setting of decentralized control over a network with
%limited bandwidth, our underlying assumption in this paper is that
%controller processing power is limited, and hence that the problem
%to be solved by the controller must be simplified.
%In fact, if the
%control computations are distributed among several processors (one
%per input, typically) then MMPC requires the network to have a
%sufficiently high bandwidth for each processor's decisions to be
%transmitted to the next processor essentially instantaneously.
%%%  END OF JMM ADDITIONS 13.04.07:
%%% Arthur's suggestion, 1 Aug 08
We emphasize that the driving factor behind the development of MMPC
is operation in a processor-limited environment, motivating
decomposition of the optimisation to reduce computational delays.
Therefore we have not considered the impact of communication limits:
indeed, in many applications of MMPC, the computation may take place
serially on a single processor, and thus communication is not a
concern.
%%% end Arthur's suggest, 1 Aug 08

%As far as we are aware, the original

%The key feature of the scheme proposed in this paper is that the
%inputs are updated sequentially, and that each control update takes
%account of all the information available at that time, namely
%knowledge of all updates already performed, and of the latest sensor
%outputs.

%Readers familiar with iterative methods of solving linear
%equations may find the following comparison helpful: the distinction
%between previous proposals for distributed MPC and our proposal for
%MMPC is analogous to the distinction between the Jacobi and the
%Gauss-Seidel iterative algorithms~\cite{BarEtal:94}. The Jacobi
%algorithm updates every variable using values only from the previous
%iteration.
%% the order in which the variables are solved is irrelevant.
%The Gauss-Seidel algorithm, on the other hand, immediately uses new values of those variables that
%have already been updated within the current iteration.
%%and so the order in which the variables are solved matters.
%Multiplexed MPC shares this idea with the Gauss-Seidel method; the move for the current actuator
%takes into account moves already made by other actuators in the same cycle of iterations.
%for  in
%the sense that it solves the control moves for each channel in the multivariable plant in an
%ordered and cyclic manner. However, Multiplexed MPC scheme is distinct in that it could also
%incorporate constraints.

In \cite{RosSheCheSha05} MPC is considered with opposite assumptions
to ours on update rates. There the plant inputs are considered to be
updated relatively frequently, compared with the rate at which
output measurements become available. This is in contrast to MMPC,
in which the plant outputs are assumed to be measured relatively
frequently, compared with the rate at which inputs are updated. It
is remarked in \cite{RosSheCheSha05} that the predictive control law
which results (with the specific assumptions made there) is
periodic, the period being the ratio of the input update rate to the
output measurement rate (assuming this is an integer). A similar
observation is central to the development in our section
\ref{stabilitymmpc}.

An alternative strategy for speeding up the computations involved in
MPC is `explicit MPC', which involves off-line precomputation of the
`pieces' of the piecewise-affine controller which is the optimal
solution~\cite{MoBaBo:03}. But that is not feasible if the number of
`pieces' required is excessively large, or if the constraints or the
plant model change relatively frequently.

MMPC was introduced by us in~\cite{LinMacWu05}. Robust MMPC was
first described in~\cite{RicLinMac07}. In \cite{RicSinLit08} our
MMPC idea was applied (by others) to the control of an aircraft
engine. In \cite{LinHoWu09} an experimental evaluation of MMPC is
reported.

\subsection{Structure of the paper}

The rest of this paper is organized as follows.
%\fbox{REWRITE THIS}\\
In Section~\ref{formulation} %two possible schemes for
a formulation of MMPC is presented in detail.
Section~\ref{stabilitymmpc} establishes the nominal stability of
MMPC with this formulation. Section \ref{sec:MMPC-cost} then derives
a formula for the value of the cost function attained by MMPC.
%Section5 analyzes the analogousness of MMPC and decentralized MPC to the
%iterative algorithms to inverting a matrix, it also demonstrate that
%MMPC can be employed to solve this kind of linear algebra problem
%very efficiently under some situation, this can be very instructive
%as matrix inversion takes up most computation resources in the
%application of `MPC on a chip`.
%%KVL010706
%An unexpected off-shoot of this research is that it provided us a system-theoretic way of looking
%at the Gauss-Seidel algorithm for solving a system of linear equations $Mz=b$. We present this in
%Section~\ref{Mxb}, and propose an alternative iterative algorithm with guaranteed convergence
%properties. This alternative algorithm is just a special case of MMPC.
 Section \ref{robustmmpc} develops a more elaborate
 formulation of MMPC, with the objective of guaranteeing robust feasibility,
 and establishes an appropriate theorem.
 Section~\ref{example} gives numerical simulation examples %which
 and compare the performance of
MMPC with SMPC
%in both the nominal case, and
for cases with
significant plant uncertainty, represented by unknown but bounded
disturbances. Finally, concluding remarks are given in
Section~\ref{conclusion}.
%For completeness, we include, in the
%appendix, derivation of the equivalent LQ problem for SMPC.
%For completeness, we
%include, in the appendix, derivations of the formula to compute
%the terminal costs.
%weight and the stabilizing gain matrices for MMPC.

%Section 5 presents simulation examples to demonstrate the
%performance benefits of the proposed scheme. Finally, concluding
%remarks are given in Section 6.

% Edited by JMM, 10.08.06.
% Edited by KVL, 13.04.07
% Edited by JMM, 14.07.08.
% Edited by JMM, 15.07.08.
% Edited by KVL, 15.08.08
% Edited by JMM, 27.07.08.
% Edited by JMM, 18.09.08 and 14.10.08.
% Edited by JMM, 29.07.09. (Contents of "structuredmmpcNEW.tex" in now inline.)
% Edited by JMM, 10.08.09 and 22.08.09 and 26.08.09.
% Edited by JMM, 20.12.09.
% Edited by KVL, 04.02.10.

\section{Problem formulation} \label{formulation}

\subsection{Preliminary}

We consider the following discrete-time linear plant model in
state-space form, with state vector $x_k\in \mathbb{R}^n$ and $m$
(scalar) inputs $u_{1,k},\ldots,u_{m,k}$:
\begin{align} \label{sysA}
 x_{k+1} = Ax_k + \sum_{j=1}^m B_j\du_{j,k}
\end{align}
where each $B_j$ is a column vector and
$\du_{j,k}=u_{j,k}-u_{j,k-1}$. (This could be generalized to the
case where $B_j\in \mathbb{R}^{n\times p_j}$ and $\du_{j,k}\in
\mathbb{R}^{p_j}$, with $\sum_j p_j$ inputs.) We assume that $(A,
[B_1,\ldots,B_m])$ is stabilizable. For ease of notation, when we
drop the index $j$, we mean the complete $B$ matrix and the input
vector so that the system (\ref{sysA}) may be written as
\[  x_{k+1} = Ax_k + B\du_{k}
\]

%We wish to devise a control strategy based on MPC which, at
%discrete-time index $k$, changes only plant input
%$(k~\text{mod}~m)+1$. In this paper, we consider two alternative
%schemes for determining the appropriate plant inputs. In both
%schemes, an increase of $k$ by 1 corresponds to a time duration of
%$T/m$, where $T$ is the complete update cycle duration
%--- see section \ref{intro}.

%move this para. to the stability section
%In both schemes an infinite prediction horizon has been chosen,
%because that is one way of ensuring closed-loop stability with
%MPC~\cite{Mac:02}. Alternative ways of obtaining stability exist,
%such as combining a finite horizon with a terminal constraint or a
%suitable terminal weight in the cost function --- see
%\cite{MayRawRaoSco:00} for a comprehensive survey.

We assume %in both schemes
that at time step $k$ the complete state vector $x_k$ is known
exactly from measurements. We will consider only the regulation
problem in detail, but tracking problems, especially those with
non-zero constant references, can be easily transformed into
equivalent regulation problems~\cite[sec.3.3]{BitGevWer90}.
%If the reference vector is constant, all this would
%require is a shift from the origin in both the state and the input
%vector to their steady state values.

\begin{ignore} %KV, 24April08
As we will be referring to the expression $(k~\text{mod}~m)+1$ often
in this paper, it is convenient to introduce the indexing function
\begin{equation} \label{eq:sigmak}
  \sigma(k) = (k~\text{mod}~m)+1
\end{equation}
The constraint
\begin{equation} \label{duconstraint1}
 \du_{j,k+i}=0 ~\text{if}~
 j\neq \sigma(k+i)
\end{equation}
then expresses our desired control updating pattern as shown in
Fig.~\ref{fig:inputpattern}.

An alternative representation of the system described by
(\ref{sysA}), together with the desired control updating pattern
(\ref{duconstraint1}) is as a periodic linear system with one input:
\begin{equation} \label{eq:periodicplant}
x_{k+1} = Ax_k + {B}_{\sigma(k)}\Delta\tilde{u}_k
\end{equation}
where $\Delta\tilde{u}_k=\du_{\sigma(k),k}$.
%Here we assume that the periodic pair $(A,B_{(\cdot)})$ is sterilizable.

From this point onwards, we use this description of the plant and we
exploit known results on periodic systems.
\end{ignore}

Multiplexed MPC, at discrete-time index $k$, changes only plant
input $\du_{\sigma(k),k}$, where $\sigma(k)$ is an indexing function
which identifies the input channel to be moved at each step, and is
defined as:
\begin{equation}
\sigma(k) = (k \bmod m)+1
\end{equation}
(We assume, without loss of generality, that we update input 1 at
time index 0.) The asynchronous nature of the multiplexed control
moves, as illustrated in Fig.~\ref{fig:inputpattern}, is captured by
the constraint
\begin{equation}
\Delta u_{j,k} = 0 \ \mathrm{if} \ j \neq \sigma(k).
\end{equation}
It is then possible to rewrite the system dynamics~(\ref{sysA}) as a
linear periodically time-varying single-input system:
\begin{equation} \label{eq:periodicplant}
x_{k+1} = Ax_k + B_{\sigma(k)} \Delta\tilde{u}_k
\end{equation}
where~$\Delta\tilde{u}_k = \Delta u_{\sigma(k),k}$.
%This form will be used for predictions in the optimisations and illustrates how
%MMPC can draw on results for periodic time-varying systems.
From this point onwards, we use this periodic description of the
plant so that we can draw on known results for periodic time-varying
systems.

 %% Next para moved here from end of sec.3, and changed. JMM, 14.10.08.
\begin{remark}
Some of the generalizations to which we alluded in section
\ref{basicidea} could be treated by redefining the sequencing
function $\sigma(\cdot)$ appropriately. For example, for a
particular 3-input system, updating the inputs in the sequence
$(1,2,1,3,1,2,1,3,\ldots)$, thus updating one of the inputs twice as
often as the others, could be represented in this way.
\end{remark}

The unique advantage of MPC, compared with other control strategies,
is its capacity to take account of constraints in a systematic
manner. As usual in MPC, we will suppose that constraints may exist
%on the input amplitudes, $\|u_k\|_\infty\leq U$,
%$\|u_k\|_\infty\leq{\mathbb U}_k$,
on the input moves, $\du_k \in \mathbb{U}_{\sigma(k)}$,
%$\|\du_k\|_\infty\leq{\mathbb U}_k$,
and on states, $x_k\in\mathbb{X}$, where $\mathbb{X}$ and
$\mathbb{U}_{\sigma(k)}$ are compact polyhedral sets containing the
origin in their interior.  %$Mx_k\leq{\mathbb X}$.
Note that the control move set depends on the time, since the
channel to be moved differs from step to step.  If constraints on
the actual control inputs $u$ are required, then $u$ must appear in
the augmented state $x$, and those constraints can be incorporated
in the state constraint set $\mathbb{X}$.
%We shall later write
%$\|u_k\|_\infty\leq{\mathbb U}_k$, $\|\du_k\|_\infty\leq{\mathbb
%U}_k$, and $Mx_k\leq{\mathbb X}$, where the sets ${\mathbb U}_k$ and
%${\mathbb X}$ will be assumed to be polytopic. The input constraint
%set ${\mathbb U}_k$ depends on $k$ in general, because input
%constraints may arise from state constraints, and hence may depend
%on $x_k$ in general.
%With this in mind, we assume
%that the states and inputs of system (\ref{sysA}) are subject to
%constraints
%\begin{equation} \label{eq:constraints}
%  x_k \in \mathbb{X} \subseteq \mathbb{R}^n, \quad
%  u_{k-1} \in \mathbb{U} \subseteq \mathbb{R}^m, \quad
%  \forall k\ge 1
%\end{equation}
%where $\mathbb{X}$ and $\mathbb{U}$ are compact polyhedral sets
%containing the origin in their interior.

Let $N=(N_u-1)m+1$ where $N_u$ is the control horizon, a design
parameter which will later be used to denote the number of control
moves to be optimized
%considered in MPC optimisation
{\em per input channel} of the original system \eqref{sysA}. The
$N$-step prediction model at time $k$ for the system described by
\eqref{eq:periodicplant} is
\begin{equation} \label{eq:prediction1}
    \vec{X}_{k+1|k} = \Phi x_{k|k} + G_{\sigma(k)}\Delta\vec{U}_{k|k}
\end{equation}
where
\[ \vec{X}_{k+1|k} = \left[\begin{array}{c}
    x_{k+1|k}\\ x_{k+2|k}\\ \vdots\\ x_{k+N|k}
    \end{array}\right], \qquad
   \Delta\vec{U}_{k|k} = \left[\begin{array}{c}
    \dtilu_{k|k}\\ \dtilu_{k+1|k}\\ \vdots\\ \dtilu_{k+N-1|k}
    \end{array}\right], \qquad
   \Phi = \left[\begin{array}{c}
    A\\ A^2\\ \vdots\\ A^N
    \end{array}\right],
\]
\begin{equation} \label{eq:Gsigmak}
  G_{\sigma(k)} = \left[\begin{array}{cccc}
    B_{\sigma(k)}        & 0               & \ldots & 0 \\
    AB_{\sigma(k)}       & B_{\sigma(k+1)} & \ldots & 0 \\
    \vdots               &                 & \ddots \\
    A^{N-1}B_{\sigma(k)} & \ldots          & AB_{\sigma(k+N-2)} &
                                                   B_{\sigma(k+N-1})
    \end{array}\right]
\end{equation}
$\Delta\tilde{u}_{k+i|k}$ denotes the prediction made at time $k$ of
a control move to be executed at time $k+i$, and $x_{k+i|k}$ denotes
the corresponding prediction of $x_{k+i}$, made at time $k$.

%\begin{remark} Different assumptions will lead to different MMPC scheme and perhaps decentralised
%scheme.
%\end{remark}
%%% End of included file "structuredmmpcNEW.tex

%%KVL010706
%Before going further, let us introduce the following definition.
%\begin{definition}
%We will use $\mathcal{X}_I(K(\cdot))$ to denote the maximum positively invariant set for the linear
%periodic system (\ref{eq:periodicplant}), when a stabilizing linear periodic feedback controller
%$K(\cdot)$ is applied, and the constraints (\ref{eq:constraints}) are satisfied~\cite{Bla99}.

%\begin{equation*}
%  \begin{array}{lllll}
%    \mathcal{X}_I(K(\cdot)) &=&
%        \{ x_0\in\mathbb{R}^n &|& x_k\in\mathbb{X},\
%        K(\cdot)x_k\in\mathbb{U}, \\
%        &&&& x_{k+1} = (A-B_{\sigma(k)}K(\cdot))x_k,\ \forall k\ge 0
%        \}
%  \end{array}
%\end{equation*}

%\end{definition}

%\begin{definition}
%The $N$-step feasible set $\mathcal{X}_N\subseteq\mathbb{R}^n$ is
%the set of states $x_k$ for which the optimal control problem
%(\ref{eq:optimalcontrol1}) is feasible, i.e.
%
%\end{definition}

\ifTA
\begin{TA} One contribution of this paper is to establish the
link between periodic linear systems and the various
decentralised/coordinated/multi-rate/multiplexed schemes of MPC.
See (Meyer and Burrus, IEEE T-CAS, 1975).
\end{TA}
\else \fi

\subsection{The MMPC Algorithm} \label{MMPCalgorithm}

%Before presenting the MMPC algorithm, we introduce the following
%notations.

%Let $\mathbb{X}$ and $\mathbb{U}_{\sigma(k)}$ be compact polyhedral
%sets containing the origin in their interior. These sets correspond
%to admissible values of states and control moves, respectively.
% $\mathbb{U}_k$ is allowed to depend on $k$ so as to allow
% constraints on $\tilde{u}_k$ as well as on $\Delta\tilde{u}_k$.
%%% Change suggested by Arthur, 01.08.08
%Note that the control move set depends on the time, since the
%channel to be moved differs from step to step.  If constraints on
%the actual control inputs $u$ are required, then $u$ must appear in
%the augmented state $x$, and those constraints can be incorporated
%in the state constraint set $\mathbb{X}$.
%%% end Arthur's suggestion

In the following, $K_{\sigma(k)}$ denotes a pre-specified
stabilizing linear periodic state-feedback
 controller of \eqref{eq:periodicplant};
$\left(\mathcal{X}_I(K_{\sigma(k)})\right)$ denotes a sequence of
sets in which none of the constraints is active, and which satisfies
the `periodic invariance' condition for the linear periodic system
\eqref{eq:periodicplant} when the feedback controller
\begin{equation} \label{eq:Kx_new}
  \Delta\tilde{u}_k = -K_{\sigma(k)}x_k
\end{equation}
is applied, namely
\begin{equation*}
  x_k \in \mathcal{X}_I(K_{\sigma(k)}) \Rightarrow
    -K_{\sigma(k)}x_k \in \mathbb{U}_{\sigma(k)} \mbox{ and }
    (A-B_{\sigma(k)}K_{\sigma(k)})x_k \in
    \mathcal{X}_I(K_{\sigma(k+1)})
\end{equation*}
and of course $\mathcal{X}_I(K_{\sigma(k)})\subseteq\mathbb{X}$, for
$\sigma(k)=1,\ldots,m$.

%(\ref{eq:constraints})
%are satisfied.
%The cost function is
%\begin{equation} \label{eq:Jknew}
%  J_k = F(x_{k+N}) +
%  \sum_{i=0}^{N-1}\left(\|x_{k+i+1}\|^2_q +
%    \|\Delta\tilde{u}_{k+i}\|^2_r\right)
%\end{equation}
%where $q=q^T>0$, $r=r^T\ge0$\footnote{Check conditions for
%existence of a stabilizing LQ control.} and $F(x_{k+N})$ is a
%suitably chosen terminal cost.
%\footnote{$F(x_{k+mN})=\|x_{k+mN}\|_{P_{k+mN}}^2$ and
%$P_{k+mN}=P_{k+mN}^T>0$}.

Some assumptions must be made %by controller $p$
about those inputs %$\Delta\vec{u}_{k,i|k}, (i=1,\ldots,m-1)$
which have already been planned %by the other controllers,
 but which have not yet been executed. We will assume that all
 such planned decisions are known to the controller, and that it
 assumes that they will be executed as planned, i.e.,
\begin{equation}
   \dtilu_{k+i|k} = \dtilu_{k+i|k-1},\quad i\ne 0,m,2m,\ldots
\end{equation}
%\begin{equation} \label{eq:mmpc2constraints}
%\Delta\vec{u}_{k,i|k} = \Delta\vec{u}_{k-1,i+1|k-1}, \quad
%  (i=1,\ldots,m-1)
%\end{equation}
(In fact, new decisions will be made at time $k+i$ in the light of
new measurements.)

MMPC solves the following finite-time constrained linear periodic
control problem:
%we set up $m$
%subproblems and define the $p$-th subproblem as
\begin{equation} \label{eq:mmpc2a}
  \begin{array}{llll}
  \mathcal{P}_{\sigma(k)}(x_k):
    & \mbox{Minimise } &
      J_k = F_{\sigma(k)}(x_{k+N|k}) + \sum_{i=0}^{N-1}\left(\|x_{k+i|k}\|^2_q +
    \|\Delta\tilde{u}_{k+i|k}\|^2_r\right) \\
    & \mbox{wrt } & \dtilu_{k+i|k},\quad (i=0,m,2m,\ldots,N-1)\\
    & \mbox{s.t.\ } & \dtilu_{k+i|k}\in\mathbb{U}_{\sigma(k+i)}, \quad (i=0,\ldots,N-1) \\
  && x_{k+i|k}\in\mathbb{X}, \quad (i=1,\ldots,N-1) \\
  && x_{k+N|k}\in\mathcal{X}_I(K_{\sigma(k)}) \\
%  && -K_{\sigma(k+N)}x_{k+N|k}\in \mathbb{U}_{\sigma(k+N)}\\
  && x_{k+i+1|k} = Ax_{k+i|k} + {B}_{\sigma(k+i)}\dtilu_{k+i|k} \\%, \quad (i=0,1,\ldots,N-1) \\
  && \dtilu_{k+i|k} = \dtilu_{k+i|k-1},\quad (i\ne 0,m,\ldots,N-1)\\
%Delta\vec{u}_{k,i|k} = \Delta\vec{u}_{k,i|k-1}, \quad
%  (i=1,\ldots,m-1)  %%% this is not correct
%  && \Delta\vec{u}_{k,i|k} = \Delta\vec{u}_{k-1,i+1|k-1}, \quad
%  (i=1,\ldots,m-1)\\
%  &&\mbox{(since we assume}\\
%&&\Delta \tilde{u}_{k+i|k}=\Delta \tilde{u}_{k+i|k-1}, \forall i\neq jm)
% wubf edit end here in 09/07/08
%  && \text{assumptions about }\Delta\vec{u}_{k,i}, (i=1,\ldots,m-1)\text{ are satisfied.}
  \end{array}
\end{equation}
where $F_{\sigma(k)}(x_{k+N|k})\geq 0$ is a suitably chosen terminal
cost.

We denote the resulting optimizing control sequence as
$\Delta\mathbf{u}^o(x_k)$. % and the optimal cost $J_k^o(x_k)$.
Only the first control $\Delta\tilde{u}^o_{k}$ in
$\Delta\mathbf{u}^o(x_k)$ is applied to the system at time $k$, so
that we apply the predictive control in the usual receding-horizon
manner.

In MMPC, there are essentially $m$ MPC controllers, operating in
sequence, in a cyclic manner. They share information, however, in
the sense that the complete plant state is available to each
controller --- although not at the same times --- and the currently
planned future moves of each controller are also available to all
the others.

For clarity, we set out the following algorithm which defines
`nominal' MMPC (as contrasted with `robust' MMPC which will be
introduced in section \ref{robustmmpc}):
 \begin{algorithm}[Nominal MMPC] \label{alg:MMPC}
\begin{enumerate}
\item[]
\item \label{step:loop1} Set $k := k_0$. Initialise by solving
problem~\eqref{eq:mmpc2a}, but optimising over all the variables
$\Delta\tilde{u}_{k+i|k}, i=0,1,\ldots,N-1$.
\item \label{step:loop2} Apply control move $\Delta u_{\sigma(k),k} = \Delta \tilde u_{k|k}$
\item Store planned moves $\Delta\vec{u}_{k,m|k}$.
\item Pause for one time step, increment $k$, obtain new measurement
$x_k$.
\item Solve problem~\eqref{eq:mmpc2a}.
\item  Go to step~\ref{step:loop2}.
\end{enumerate}
 \end{algorithm}
Note that Step~\ref{step:loop1} involves solving for inputs across
all channels, not just channel~$\sigma(k)$.  This type of
initialisation requirement is common in distributed MPC.  Subsequent
results do not depend on the optimality of this initial solution,
only its feasibility.

%The
%terminal time $k+mN$ increases with time $k$ and is often referred
%to as a {\it receding horizon}.
%Note that the cost is accumulated over a prediction horizon of $N$ \emph{cycles}, that is, over
%$mN$ steps.

%where the cost function is
%\begin{equation}
%  J_k = F(x_{k+N})+
%     \sum_{i=0}^{N-1} \left(\| x_{k+i+1} \|_q^2 +
%       \| \dtilu_{k+i} \|_r^2\right)
%\end{equation}

%Note the subtle difference in the terminal time in the Scheme~2 cost function compared with that of
%Scheme~1. This is so that in Scheme~2 there will be $N$ unknown variables in each of the
%subproblems, whereas in Scheme~1 there are $mN$ unknown variables.
%This construction of the cost function is also influenced by our
%proposed MMPC implementation method for Scheme~2(see below) and
%the stability proof (see Section~\ref{stabilityscheme2}).

%The set of future input moves $\mathbf{u}^{\ne p}$ are assumed
%known from the outcome of the earlier $(m-1)$ subproblems
%$\{\mathcal{P}_N^j(x_{k-j}),\ j=1,\ldots,m.,\ \sigma(k-j)\ne p
%\}$.

\begin{ignore}
Different assumptions are possible here. We will assume that each
input is computed so as to optimize ``its'' cost over the prediction
horizon, and that after the end of the horizon each input is
determined according to an optimal linear state-feedback law.
\end{ignore}

\ifTA
\begin{TA}
  In IFCA2005 paper, we said that: If
the length of the planning horizon is the shortest possible,
namely $N=1$, then Schemes~1 and 2 are equivalent. This is not
correct. It is correct for a horizon of 1 step, but $N=1$ means a
horizon of 1 cycle,namely $m$ steps.
\end{TA}
\else \fi

%%%%%%%%%%%%%%%%%%%%%%%%%%%%%%%%%%%%%%%%%%%%%%
\section{Stability of MMPC} \label{stabilitymmpc}

%\subsection{Unconstrained LQ optimal control of periodic systems}
%\label{LQperiodic}

%%%%% Added by JMM 27.08.08:
In this section we establish sufficient conditions under which the
MMPC scheme gives closed-loop stability. We then apply standard
results on optimal control of periodic systems to our plant written
in the form of \eqref{eq:periodicplant}, assuming that all
constraints are inactive, to propose a terminal cost $F(\cdot)$
which, when used in the MMPC algorithm introduced in section
\ref{MMPCalgorithm}, ensures stability of the closed loop even when
constraints are active.
%%%%% End of JMM additions 27.08.08.

%\subsection{Stability of MMPC schemes} \label{stabilityschemes}

%\begin{proposition}
\begin{theorem} \label{thm:nominalMMPC}
MMPC, obtained by implementing the nominal MMPC Algorithm
\ref{alg:MMPC},
%solving online the finite-time constrained linear
%periodic optimal control problem %s \eqref{eq:mmpc1a} and
%\eqref{eq:mmpc2a}, %respectively,
gives closed-loop stability if the problems are well-posed, and if
the set of terminal costs $\{F_\sigma(\cdot)\}$ satisfies
\begin{equation}\label{eq:FCLF}
    F_{\sigma^+}([A-B_{\sigma^+} K_{\sigma^+}]x)+\|x\|^2_q
    +\|K_{\sigma^+}
    x\|_r^2 \leq F_\sigma(x) \quad{\rm for}\quad \sigma=1,\ldots,m.
\end{equation}
where $\sigma^+=(\sigma \text{ mod }m)+1$, namely the cyclical
successor value to $\sigma$.

%is the value function of the unconstrained infinite
%horizon periodic optimal control problem with initial condition
%$x_{k+N|k}$, namely if
%\begin{multline} \label{eq:FxkN}
%  %\begin{array}{llll}
%%    F(x_{k+N}) &=& \min_{\Delta\tilde{u}_{k+i|k}}\{
%%  \sum_{i=N}^{\infty}\left(\|x_{k+i+1|k}\|^2_q +
%%    \|\Delta\tilde{u}_{k+i|k}\|^2_r\right)\ |\
%%        x_{k+i+1|k} = Ax_{k+i|k} + B_{\sigma(k+i)}\Delta\tilde{u}_{k+i|k} \} \\
%%     &=&   x_{k+N|k}^T\bar{P}_{\sigma(k+N)}x_{k+N|k} \label{eq:xPx}
%%  \end{array}
%F(x_{k+N|k}) = \min_{\Delta\tilde{u}_{k+i|k}}\left\{
%\sum_{i=N}^{\infty}\left(\|x_{k+i+1|k}\|^2_q +
%    \|\Delta\tilde{u}_{k+i|k}\|^2_r\right)\right\}\\
%    = x_{k+N|k}^T\bar{P}_{\sigma(k+N)}x_{k+N|k}
%\end{multline}
%where $\bar{P}_{\sigma(k+N)}$ is defined as above.
\end{theorem}

{\bf Proof:}\\
The proof follows a standard argument for MPC stability proofs (see
\cite{MayRawRaoSco:00} for example), adapted to our setting. It
depends on the constrained optimization being feasible at each step,
and the feasibility at any particular time step depends on the
details of the constrained optimization problem that is being
solved. For the nominal case with a perfect model and in the absence
of disturbances, if feasible solutions are obtained over an initial
period, then feasibility is assured thereafter.

Let $\dtilu^*_{k+i|k}$ denote the optimal solution to
\eqref{eq:mmpc2a} at time step $k$, for $i=0,m,2m,\ldots,N-1$, let
$x^*_{k+i|k}$ denote the corresponding state sequence, for
$i=1,2,\ldots,N$, and let $J^*_k$ be the corresponding value of the
cost function $J_{k}$.

%%%%%%%%%%%%%%%%%%%%%%%%%%%%%%%%%%
% New proof begun by JMM, 17.09.09:

Then a candidate input sequence to be applied to the plant at the
next time step, $k+1$, is
\begin{multline}\label{eq:candsoln}
    \left(
    \dtilu^*_{k+1|k+1-m},\dtilu^*_{k+2|k+2-m},\ldots,\dtilu^*_{k+m|k},\right. \\
    \dtilu^*_{k+1+m|k+1-m},\dtilu^*_{k+2+m|k+2-m},\ldots,\dtilu^*_{k+2m|k},
    \ldots,\\
    \dtilu^*_{k+1+(N_u-2)m|k+1-m},\dtilu^*_{k+2+(N_u-2)m|k+2-m},\ldots,\\
    \left. \dtilu^*_{k+(N_u-1)m|k},
    -K_{\sigma(k+1)}x^*_{k+N|k}
    \right)
\end{multline}
(recall that $N-1=(N_u-1)m$). The input sequence applied at time
step $k$ is the same, except that the initial term $\dtilu^*_{k|k}$
is pre-pended to it, and that the final term
$K_{\sigma(k+1)}x^*_{k+N|k}$ is omitted. Let the cost obtained with
the candidate solution \eqref{eq:candsoln} be $\tilde{J}_{k+1}$.
Then
\begin{multline}\label{eq:Jdiff}
    \tilde{J}_{k+1}-J^*_k = ||K_{\sigma(k+1)}x^*_{k+N|k}||^2_r +
    F_{\sigma(k+1)}\left(
    (A-B_{\sigma(k+1)}K_{\sigma(k+1)})x^*_{k+N|k}\right) \\
    -||\dtilu^*_{k|k}||^2_r -
    ||x^*_{k|k}||^2_q - F_{\sigma(k)}(x_{k+N|k}) + \|x_{k+N|k}\|^2_q
\end{multline}
Hence $\tilde{J}_{k+1}-J^*_k\leq 0$ if \eqref{eq:FCLF} holds. Now
optimisation at time step $k+1$ will result in a value function
\begin{equation}\label{eq:JstarJinequality}
    J^*_{k+1} \leq \tilde{J}_{k+1}
\end{equation}
and hence
\begin{equation}\label{eq:JlyapProperty}
    J^*_{k+1} \leq J^*_k
\end{equation}
if \eqref{eq:FCLF} holds.

But $J_k^*\geq 0$ for all $k$, hence $J^*_{k+1} - J^*_k \rightarrow
0$. But, from \eqref{eq:Jdiff}--\eqref{eq:JlyapProperty} we have
that
\begin{equation}\label{eq:Jstardiffinequality}
    J^*_{k+1}-J^*_k \leq -\|x^*_{k|k}\|^2_q-\|\Delta u^*_{k|k}\|^2_r
\end{equation}
Hence $x^*_{k|k}\rightarrow 0$ (and $\Delta u^*_{k|k}\rightarrow
0$). But $x_{k|k}=x_k$, so $x_k\rightarrow 0$.
%Hence $J^*_k$ is a Lyapunov function for the controlled system.

%%%%%%%%%%%%%%%%%%%%%%%%%%%%%%%%%%

%By Bellman's optimality
%principle we have that $x^*_{k+N+1|k+1}=x^*_{k+N+1|k}$, and hence we
%have
%\begin{equation}
%  F(x_{k+N|k}) = F(x_{k+N+1|k+1})
%    + \|x^*_{k+N+1|k}\|_q^2 + \|\dtilu^*_{k+N|k}\|_r^2
%\end{equation}
%%with $x_{k+mN+1}=(A-B_{sigma(k+mN)}K(\cdot))x_{k+mN}$.
%from which
%\begin{equation}
%  F(x_{k+N+1|k+1}) - F(x_{k+N|k}) =
%    - \|x^*_{k+N+1|k}\|_q^2 - \|\dtilu^*_{k+N|k}\|_r^2 \le 0
%\end{equation}
%follows, with equality satisfied if and only if
%$x_{k+N|k}=\dtilu_{k+N|k}=0$. This shows that $F(x_{k+N|k})$ is a
%Lyapunov function in some neighbourhood of 0, in particular for
%$x_k\in\mathcal{X}_I(K_{\sigma(k)})$.
\hfill $\blacksquare$

\begin{remark}
Note the implicit assumption that $N$ is chosen sufficiently large
%for constraints to be inactive after the end of the finite
%optimisation horizon, namely at times $k+i, i>N$.
to ensure feasibility of the constrained optimisation problem posed.
Also note the assumption in each planning optimisation that the
linear state feedback law \eqref{eq:Kx_new} is applied at
\emph{every} step after the end of the optimisation horizon.
\end{remark}

%%%%%%%%%%%%%%%%%%%%%%%%%%%%%%%%%%%

The following results on unconstrained infinite-time linear
quadratic control of periodic systems are known~\cite{BitColDen:88}.
Consider the plant \eqref{eq:periodicplant} and the quadratic cost
function
\begin{equation}\label{eq:lqinfcost}
    J_k = \sum_{i=0}^\infty \left(\|x_{k+i}\|^2_q +
    \|\Delta\tilde{u}_{k+i}\|^2_r\right)
\end{equation}
Then this cost is minimised by finding $\bar{P}_i, i=1,\ldots,m$,
the Symmetric, Periodic and Positive Semidefinite (SPPS) solution of
the following discrete-time periodic Riccati equation (DPRE)
\begin{equation} \label{eq:periodicRDE}
    P_k = A^T P_{k+1}A - A^T P_{k+1}B_{\sigma(k)}(B_{\sigma(k)}^T P_{k+1}B_{\sigma(k)}+r)^{-1}
            B_{\sigma(k)}^T P_{k+1}A+q
\end{equation}
and setting
\begin{equation} %\label{eq:Kx}
    \dtilu_{k} = -K_{\sigma(k)} x_{k}
\end{equation}
where
\begin{equation}\label{eq:LQK}
K_{\sigma(k)} = (B_{\sigma(k)}^T
\bar{P}_{\sigma(k+1)}B_{\sigma(k)}+r)^{-1}B_{\sigma(k)}^T
   \bar{P}_{\sigma(k+1)}A
\end{equation}
Furthermore, the resulting minimal value of $J_k$ is given by
$J_k^*= x_k^T\bar{P}_{\sigma(k)}x_k$. Thus one way of choosing a
suitable set of terminal costs to satisfy \eqref{eq:FCLF} is to set
\begin{equation}\label{eq:FinfP}
    F_{\sigma(k)}(x) = x^T\bar{P}_{\sigma(k+N)}x
\end{equation}
which leads to
\begin{equation}\label{eq:Jstardiff}
    J^*_{k+1}-J^*_k = -\|x^*_{k|k}\|^2_q-\|\Delta u^*_{k|k}\|^2_r
\end{equation}

%which satisfies \eqref{eq:FCLF} with equality.

\begin{remark}
The terminal cost \eqref{eq:FinfP} would be the optimal cost-to-go
if at each step $k$ the optimisation was over future values of
\emph{all} input channels, rather than those in input channel
$\sigma(k)$ only. A version of MMPC in which this is done was called
`scheme 1 MMPC' in our earlier paper~\cite{LinMacWu05}. The version
presented in this paper was called `scheme 2' in~\cite{LinMacWu05}.
We no longer advocate `scheme 1', as it does not give any reduction
of computational complexity, compared with conventional SMPC.
\end{remark}

\begin{ignore}
Note that the terminal cost used in scheme 2 is the same as that
used in scheme 1. However, the value functions of the two problems
$\mathcal{P}$ and $\mathcal{P}_{\sigma(k)}$ are different, because
they have different sets of decision variables.
\end{ignore}

\begin{ignore}
It can be seen that this result, and its proof, depend only on using
an appropriate terminal cost $F(\cdot)$, and not at all on the
details of the constrained optimization over the horizon of length
$N$. Consequently the result and its proof hold for any other MMPC
scheme which involves constrained optimization over a finite
horizon, providing that the terminal cost is given by
\eqref{eq:FxkN}. For example, for a particular 3-input system, it
may be beneficial to update the inputs in the sequence
$(1,2,1,3,1,2,1,3,\ldots)$, thus updating one of the inputs twice as
often as the others. The stability of such a scheme is proved by the
argument used above, providing that the sequencing function
$\sigma(\cdot)$ is redefined appropriately.
\end{ignore}

\begin{ignore}
Scheme 1, if applied over a finite horizon, has a lower
computational complexity than conventional multivariable MPC,
providing that the number of steps in the horizon remains the same
--- which means, since the step duration has been reduced to
$T/m$, that a shorter time horizon is used.

Scheme 2 has a lower computational complexity than conventional
multivariable MPC, even if the same time horizon length is
retained, that is even if the number of steps is increased by a
factor $m$, since the complexity per step is reduced by $O(m^3)$.
\end{ignore}

%\begin{remark}
%If no disturbance is present, Scheme~1 and Scheme~2 will give
%identical result, since Scheme~2 is initialized using Scheme~1.
%\end{remark}

\begin{ignore} The trajectory obtained by the Scheme 2 controller will
usually be different from that found by the Scheme 1 controller. It
will be `less optimal' in the sense that it will usually result in
larger values of the cost (although each constituent controller will
be optimizing the same cost function with respect to the degrees of
freedom available to it).

{\bf Is the above still true? Perhaps with no disturbance present,
Scheme~1 and Scheme~2 will give identical result, if Scheme~2 is
initialized using Scheme~1.}

\end{ignore}

%%%%%%%%%%%%%%%%%%%%%%%%%%%%%%%%%%%%%%%%%%%%%%%%%%%%%%%%%%%%%%%
\begin{ignore} %this is the previous version
\subsection{Stability of Scheme 1} The stability of Scheme 1 with
constraints can be established as follows:

Let $J_k^o$ be the optimal value of $J_k$ achieved by controller
at time $k$.
%Stability follows from the fact that
%$J_{k+1}^o\leq J_k^o$, with equality holding only if $x_k=0$. This
%monotonic decreasing property of $J_k^o$ can be established by a
%standard argument:
Suppose that at step $k+1$, the controller chooses
$\Delta\tilde{u}_{k+mN}= -\kappa_f(\cdot)x_{k+mN}$ and leaves the
moves $\{ \Delta\tilde{u}_{k+i}, i=1,\ldots,mN-1 \}$ unchanged
from the values which were assumed by the controller at step $k$,
and that this achieves the value $J_{k+1}$. Then
\begin{equation}\label{eq:decreasingJk}
    J_{k+1} = J_k^o + F(x_{k+mN+1}) - F(x_{k+mN})
    + \|x_{k+mN}\|_q^2 + \|\Delta\tilde{u}_{k+mN}\|_r^2
    - \|x_{k}\|_q^2 - \|\Delta\tilde{u}_k\|_r^2
\end{equation}
where $x_{k+mN+1} = (A-B(\cdot)\kappa_f(\cdot))x_{k+mN}$.

If we choose $F(x_{k+mN})$ to be the value function of the
unconstrained infinite horizon (periodic) optimal control problem,
i.e.
\begin{equation}
  F(x_{k+mN}) = \sum_{i=mN}^{\infty}\left(\|x_{k+i}\|^2_q +
    \|\Delta\tilde{u}_{k+i}\|^2_r\right) =
    x_{k+mN}^TP_{k+mN}x_{k+mN}
\end{equation}

\begin{equation}
  \mathcal{P}_\infty^{\mbox{uc}}: J(x_k)_\infty^{\mbox{uc}} =
        \sum_{i=mN}^{\infty}\left(\|x_{k+i}\|^2_q +
    \|\Delta\tilde{u}_{k+i}\|^2_r\right)
\end{equation}
then $\kappa_f(\cdot)$ will be the appropriate stabilising
periodic state feedback gain.

As a consequence
\begin{equation}
F(x_{k+mN+1}) - F(x_{k+mN})
    + \|x_{k+mN}\|_q^2 + \|\Delta\tilde{u}_{k+mN}\|_r^2 = 0
\end{equation}
and %This cost, which is an upper bound for $J_{k+1}^o$, satisfies
\begin{equation}
  J_{k+1}^o \le J_{k+1} = J_k^o - \|x_{k}\|_q^2 - \|\Delta\tilde{u}_k\|_r^2
\end{equation}
This shows that $J_k^o$ is a Lyapunov function for the overall
controller of Scheme~1, and stability is established.

Here, we have implicitly assumed the terminal constraint
\begin{equation}
  x_{k+mN} \in \mathcal{X}_f
\end{equation}
where $\mathcal{X}_f$ is the output admissible set for $x_{k+m} =
\Psi x_k$ and $\Psi$ is the monodromy matrix of the resulting
unconstrained periodic closed-loop system and has all its
eigenvalues inside the unit circle. More specifically,
\begin{eqnarray}
  \mathcal{X}_f(\kappa_f(\cdot)) := \{ x(0)\in \mathcal{R}^n &|&
    x(k)\in \mathcal{X}, \kappa_f(\cdot)x(k)\in\mathcal{U}_N(x_k), \nonumber\\
    && x(k+1)=(A-B(\cdot)\kappa_f(\cdot))x(k), \forall k\ge 0\}
\end{eqnarray}

According to \cite{MayRawRaoSco:00}, with the ingredients,
$F(\cdot) = J(\cdot)_\infty^{\mbox{uc}}$ for all
$x\in\mathcal{X}_f$. the closed-loop is exponentially stable with
a domain of attraction $X_N$ (the set that can be controlled by
MMPC with fixed horizon $N$).

\section{Stability of Scheme~2} Stability of Scheme~2 cannot be
proved in exactly the same way as in Scheme~1, since the
computation of a future input trajectory is done for one subset of
inputs at a time.

The stability of Scheme~2 can be developed as follows: Let $J_k^o$
be the optimal value of $J_k$ achieved by controller $\sigma(k)$.
%Stability follows from the fact that $J_{k+1}^o\leq J_k^o$, with
%equality holding only if $x_k=0$. This monotonic decreasing
%property of $J_k^o$ can be established by a standard
%argument:
Suppose that at step $k+1$, controller $\sigma(k+1)$ leaves the
moves $\Delta\tilde{u}_{k+1},\ldots,\Delta\tilde{u}_{k+mN-1}$
unchanged from the values which were assumed by controller
$\sigma(k)$ at step $k$, and that this achieves the value
$J_{k+1}$. Then
\begin{equation}\label{eq:decreasingJk}
    J_{k+1} = J_k^o - \|x_{k+1}\|_q^2 - \|\Delta\tilde{u}_k\|_r^2
\end{equation}
But $J_{k+1}^o\leq J_{k+1}$, from which the conclusion follows.
Consequently $J_k^o$ is a Lyapunov function for the overall
controller of Scheme~2, and stability is established.

\begin{remark}
In both schemes the infinite horizon cost has been chosen as the
terminal cost, because that is one way of ensuring closed-loop
stability with MPC~\cite{Mac:02}. Alternative ways of obtaining
stability exist, such as combining a finite horizon with a
terminal constraint or a suitable terminal weight in the cost
function --- see \cite{MayRawRaoSco:00} for a comprehensive
survey.
\end{remark}

\begin{remark}
Scheme 1, if applied over a finite horizon, has a lower
computational complexity than conventional multivariable MPC,
providing that the number of steps in the horizon remains the same
--- which means, since the step duration has been reduced to
$T/m$, that a shorter time horizon is used.

Scheme 2 has a lower computational complexity than conventional
multivariable MPC, even if the same time horizon length is
retained, that is even if the number of steps is increased by a
factor $m$, since the complexity per step is reduced by $O(m^3)$.
\end{remark}

\begin{remark}
If no disturbance is present, Scheme~1 and Scheme~2 will give
identical result, since Scheme~2 is initialised using Scheme~1.
\end{remark}

\begin{ignore} The trajectory obtained by the Scheme 2 controller will
usually be different from that found by the Scheme 1 controller.
It will be `less optimal' in the sense that it will usually result
in larger values of the cost (although each constituent controller
will be optimising the same cost function with respect to the
degrees of freedom available to it).

{\bf Is the above still true? Perhaps with no disturbance present,
Scheme~1 and Scheme~2 will give identical result, if Scheme~2 is
initialised using Scheme~1.}

\end{ignore}

%%%%%%%%%%%%%%%%%%%%%%%%%%%%%%%%%%%%%%%%%%%%%%%%%%%%%%%%%%%%%%%%

%%% From ICARCV2008 paper, KVL 24 April 2008, Cambridge

% Edited by JMM, 20.12.09.
% Edited by JMM, 22.08.09, 26.08.09.
% Edited by JMM, 29.07.09. (proposition2.tex now inline in this file)
% Edited by JMM & KVL, 18.07.08, NTU.
% Edited by KVL, 15.08.08
% Edited by JMM, 27.08.08
% Edited by JMM, 18.09.08, 13.10.08.

\section{Cost of MMPC when constraints are inactive} \label{sec:MMPC-cost}

%%%%%%%%  JMM added this paragraph %%%%%%%%%%%%%%%%%%%%%%%
Each solution of the optimisation problem \eqref{eq:mmpc2a} depends
on the plans made in previous optimisations. Hence the optimal cost
obtained with MMPC, even in the case that all constraints are
inactive, is not given by \eqref{eq:FinfP}. In this section we will
introduce an augmented state which includes those existing plans
that are not going to be modified by the current optimisation. This
will allow us to obtain, in Theorem \ref{prop:MMPCcost}, an
expression for the optimal cost of the same form as
\eqref{eq:FinfP}. This will provide an analysis tool for predicting
and comparing the performance of various MMPC designs. In the
process we will see that MMPC can be rewritten in a more familiar
MPC form, but with a periodically time-varying (augmented state)
model.
%In this
%section we shall pursue that view, in order to obtain an expression
%for the cost obtained with any linear periodic stabilising
%controller, which updates the plant inputs sequentially as in MMPC.
%As a special case the cost of MMPC, assuming that constraints are
%never active, can be obtained.

Note that a similar development could be used to compute the optimal
MMPC cost if the set of active constraints was constant and known.
The nature of the MMPC control law in such circumstances is also
linear periodic. Consequently the MMPC control law in the presence
of constraints is piecewise-linear-periodic; as in the standard
`explicit' MPC case, the `pieces' correspond to regions of the state
space in which the set of active constraints remains constant.

The development of this section, in particular Theorem
\ref{prop:MMPCcost}, facilitates performance evaluation of MMPC in
certain circumstances. For example, it is useful for evaluating the
trade-off between the restricted optimisation performed by MMPC and
the reduced update rate available with conventional SMPC.

%%%%%%%%%  wubf add this paragraph %%%%%%%%%%%%%%%%%%%%%%%
%In this section, we will show that MMPC can be interpreted as a
%piecewise linear periodic state feedback law in an enlarged state
%space. This provides%, for the first time,
%a framework for systematic
%analysis and investigation of the properties of MMPC, enabling a
%comparative study between MMPC and SMPC. In particular, we derive
%formula for computing the quadratic cost of MMPC.
%%%%%%%%%%%%

\subsection{Unconstrained MMPC as periodic state feedback} \label{sec:main}

We introduce the following definitions, which gather together those
variables which are optimised at each step by the MMPC algorithm:
\begin{equation} \label{eq:delvecukik}
 \Delta\vec{u}_{k,i|k} = \left[\begin{array}{c}
    \dtilu_{k+i|k} \\ \dtilu_{k+m+i|k} \\ \vdots \\
    \dtilu_{k+(N_u-2)m+i|k}
    \end{array}\right]
\end{equation}
for $i=1,2,\ldots,m$.
 \begin{equation} \label{eq:delvecuk0k}
 \Delta\vec{u}_{0|k} = \left[\begin{array}{c}
    \dtilu_{k|k} \\ \dtilu_{k+m|k} \\ \vdots \\
    \dtilu_{k+(N_u-1)m|k}
    \end{array}\right]
  = \left[\begin{array}{c}
    \dtilu_{k|k} \\ \Delta\vec{u}_{k,m|k}
    \end{array}\right]
\end{equation}

Recall that $N=(N_u-1)m+1$ where $N_u$ is the control horizon, a
design parameter which denotes the number of control moves to be
optimized
%considered in MPC optimisation
per input channel of the original system \eqref{sysA}. By grouping
the predicted control signals into $m$ vectors, the prediction model
\eqref{eq:prediction1} can be re-written as
\begin{equation} \label{eq:regroupprediction}
 \vec{X}_{k+1|k} = \Phi x_{k|k} +
g^{\sigma(k)}_1\Delta\vec{u}_{0|k}
    + g^{\sigma(k)}_2\Delta\vec{u}_{k,1|k}
    + \ldots + g^{\sigma(k)}_m\Delta\vec{u}_{k,m-1|k}
\end{equation}
where $\Delta\vec{u}_{k,i|k}$ and $\Delta\vec{u}_{0|k}$ are as
defined in \eqref{eq:delvecukik} and \eqref{eq:delvecuk0k},
respectively, and $g^{\sigma(k)}_i,\ (i=1,\ldots,m)$ are matrices
whose columns are columns of the $G_{\sigma(k)}$ matrix
\eqref{eq:Gsigmak}, namely, $g^{\sigma(k)}_1$ is the matrix whose
columns are columns $1,1+m,\ldots,1+(N_u-1)m$ of the matrix
$G_{\sigma(k)}$, while $g^{\sigma(k)}_i, (i=2,\ldots,m)$ contains
columns $i,i+m,\ldots,i+(N_u-2)m$ columns of $G_{\sigma(k)}$.

In MMPC only $\Delta\vec{u}_{0|k}$ is taken as the decision variable
at time $k$, and appropriate assumptions are made about
$\Delta\vec{u}_{k,i|k},\ i=1,\ldots,m-1$. Note that the length of
$\Delta\vec{u}_{0|k}$ is $N_u$ while the length of
$\Delta\vec{u}_{k,i|k}$ for $i=1,\ldots,m-1$ is $N_u-1$. When
$N_u=1$, $\Delta\vec{u}_{k,i|k}$, $i=1,\ldots,m-1$, become
zero-length vectors.
%Eq.(\ref{eq:regroupprediction}) gives the prediction equation, with
%the controls grouped according to the different input channels.
%The constrained optimisation problem formulation for MMPC optimises
%with respect to one channel while making assumptions about other
%control channels.
In MMPC we assume that the $\Delta\vec{u}_{k,i|k}, i=1,\ldots,m-1$
are those inputs which have already been planned in previous steps
but have not yet been executed, namely
\begin{equation}\label{eq:dukassumption}
% \Delta\vec{u}_{k+1,i|k+1} = \Delta\vec{u}_{k,i+1|k}, \qquad
 \Delta\vec{u}_{k,i|k} = \Delta\vec{u}_{k-1,i+1|k-1}, \qquad
 (i=1,\ldots,m-1).
\end{equation}
We define the vector $\Delta \vec{u}_{k|k-1}^p$ which holds the
%``predicted control vectors from the other $(m-1)$ control channels''
previously planned but not yet executed control moves as
\begin{equation} \label{eq:delukp}
%\Delta \vec{u}_{k|k-1}^p=\left[\begin{array}{c}
%     \Delta \vec{u}_{k,1|k-1}\\
%     \Delta \vec{u}_{k,2|k-1}\\
%     \vdots\\
%     \Delta \vec{u}_{k,m-1|k-1} \end{array}\right]
\Delta \vec{u}_{k|k-1}^p
%     =\left[\begin{array}{c}
%     \Delta \vec{u}_{k,1|k}\\
%     \Delta \vec{u}_{k,2|k}\\
%     \vdots\\
%     \Delta \vec{u}_{k,m-1|k} \end{array}\right]
     =\left[\begin{array}{c}
     \Delta \vec{u}_{k-1,2|k-1}\\
     \Delta \vec{u}_{k-1,3|k-1}\\
     \vdots\\
     \Delta \vec{u}_{k-1,m|k-1} \end{array}\right]
\end{equation}
%We will show that,
Thus, it can be deduced from \eqref{eq:regroupprediction} that, if
no constraints are active,
% or the set of constraints which will be active is known,
then the MMPC control law is a linear periodic state
feedback:
%function of $x_k$ and $\Delta\vec{u}_{k,i|k-1}, i=1,2,...,m-1$.
%To this end, we %introduce the notation $\Delta\vec{u}_{k,i|k+i-m}$
%%and
%Thus, if we denote the optimal control sequence of MMPC at time step
%$k$ as $\Delta \vec{u}_k$, the MMPC control law has the form
\begin{equation}\label{eq:mmpclaw}
   \Delta \vec{u}_{0|k} = \tilde{K}_{\sigma(k)} \xi_k
\end{equation}
%i.e., at time step $k$, the control update is determined by $x_k$ as
%well as $\Delta \vec{u}_{k|k-1}^p$, the predicted control trajectory
%from the other $m-1$ control channels.
where we have introduced the augmented state vector
\begin{equation}\label{eq:xi}
    \xi_ k = \left[\begin{array}{c}
     x_k \\ \Delta \vec{u}_{k|k-1}^p\\
    \end{array}\right]
\end{equation}

%Since only the first element of the $\Delta \vec{u}_{k,0|k}$ vector
%is applied to the system at time $k$, with the remaining $N_u-1$
%elements to be used in subsequent MMPC updates,
%%we further partition $\Delta\vec{u}_{k|k}$ as
%it may be useful to note that
%\[\Delta \vec{u}_{k|k} = \left[\begin{array}{cc}
%     \dtilu_{k|k} & \Delta \vec{u}^T_{k,m|k} \end{array}\right]^T
%\]
%where
%\[ \Delta
%\vec{u}_{k,m|k}=\left[\begin{array}{c}
%     \dtilu_{k+m|k}\\
%     \dtilu_{k+2m|k}\\
%     \vdots\\
%     \dtilu_{k+(N_u-1)m|k} \end{array}\right]
%\]

\subsection{A formula for the MMPC Cost}

%From Eq.(\ref{eq:mmpclaw}), it is clear that, if $N_u=1$ and the
%constraints are not active, the MMPC problem defined in
%Eq.(\ref{eq:mmpc2a}) coincides with standard periodic control since
%in this case $\Delta \vec{u}_{k|k-1}^p$ vanishes. Thus, in this
%case, the MMPC cost coincides with the infinite horizon cost and is
%given by
%\[  J_{\sigma(k)} = \sum_{i=0}^{\infty}\left(\|x_{k+i+1|k}\|^2_q+
%\|\Delta\tilde{u}_{k+i|k}\|^2_r\right)
%  = x_{k|k}^T P_{\sigma(k)} x_{k|k}
%\]
%In general, however, when $N_u>1$, and when constraints are not
%active, the periodic control law obtained for the MMPC scheme will
%be different from that of the standard periodic state feedback law
%and the usual formula will not apply.

Using the augmented state vector introduced in \eqref{eq:xi}, the
dynamics of the plant operating under MMPC can be expressed as
\begin{equation}\label{eq:augmentplant}
  \xi_{k+1} = \tilde{A}\xi_k+\tilde{B}_{\sigma(k)}\Delta
\vec{u}_{k,0|k}
\end{equation}
where
\[
\tilde{A} =\left[%
\begin{array}{cc}
  A & 0 \\
  0 & A_u\\
\end{array}%
\right]\quad
\tilde{B}_{\sigma(k)}=\left[%
\begin{array}{cc}
  B_{\sigma(k)} & 0 \\
  0 & B_u\\
\end{array}%
\right]
\]
\[
A_u= \left[%
\begin{array}{ccccc}
  0 & I & 0 & \cdots & 0 \\
  0 & 0 & I &  & 0 \\
  \vdots &  & \ddots & \ddots & \vdots \\
  \vdots &  &  &  & \vdots \\
  0 &  & \cdots & \cdots & 0 \\
\end{array}%
\right]\qquad
B_u= \left[%
\begin{array}{c}
  0 \\
  0 \\
  \vdots \\
  \vdots \\
  I \\
\end{array}%
\right]
\]

%The cost \eqref{eq:lqinfcost}, when control law of the form
% \eqref{eq:mmpclaw} is applied, can then be written as
%Then,
%Eq.(\ref{eq:augmentplant}), together with Eq.(\ref{eq:mmpclaw}),
%enable one to re-cast the (unconstrained) MMPC formulation under the
%standard periodic control framework. More specifically, we could now
%compute the MMPC infinite horizon cost
%\[ J_{\sigma(k)} =
%\sum_{i=0}^{\infty}\left(\|x_{k+i+1}\|^2_q+
%\|\Delta\tilde{u}_{k+i}\|^2_r\right)
%\]
%for the system (Eq.(\ref{eq:periodicplant})) under the MMPC law
%(Eq.(\ref{eq:mmpclaw})). This cost can be computed indirectly
%through

The value of the quadratic cost (see also Remark~\ref{rem:4})
\begin{equation} \label{eq:Jtilde}
J_k =
\sum_{i=0}^{\infty}\left(\|\xi_{k+i+1}\|^2_{\tilde{q}}+\|\Delta
\vec{u}_{0|k+i}\|_{\tilde{r}}^2\right)
\end{equation}
when control law of the form \eqref{eq:mmpclaw} is applied, is given
by the following theorem:

%\input{proposition2} % Proof
%\begin{proposition}
%Given the stabilising MMPC control law $\tilde{K}_{\sigma(k)}$, the
%infinite horizon cost of (23) for the system (\ref{eq:augmentplant})
%is given by
%\begin{proposition} \label{prop:MMPCcost}
\begin{theorem} \label{prop:MMPCcost}
The value of the cost \eqref{eq:Jtilde} for the system
\eqref{eq:augmentplant}, when any stabilising linear periodic state
feedback of the form \eqref{eq:mmpclaw} is applied, is given by
\begin{equation} \label{eq:Jtildesum}
\tilde{J}_k = J_{\xi,k}+J_{u,k}
   = \xi_k^TP_{\xi+u,\sigma(k)}\xi_k
\end{equation}
where $J_{\xi,k} = \xi_k^TP_{\xi,\sigma(k)}\xi_k$ and $J_{u,k} =
\xi_k^TP_{u,\sigma(k)} \xi_k$, and $P_{\xi,\sigma(k)}$,
$P_{u,\sigma(k)}$ and $P_{\xi+u,\sigma(k)}$ are, respectively,
solutions of the following Lyapunov equations: {\small
\begin{eqnarray}
%\nonumber
P_{\xi,\sigma(k)}&=&\Psi_{\sigma(k)}^TP_{\xi,\sigma(k)}\Psi_{\sigma(k)}+\bar{\Phi}_{\sigma(k)}^T
Q\bar{\Phi}_{\sigma(k)} \label{eq:Pxi} \\
%\nonumber
P_{u,\sigma(k)}&=&\Psi_{\sigma(k)}^TP_{u,\sigma(k)}\Psi_{\sigma(k)}+\bar{K}_{\sigma(k)}^T
R\bar{K}_{\sigma(k)} \label{eq:Pu} \\
%\label{eq:Pmmpc}
P_{\xi+u,\sigma(k)}&=&\Psi_{\sigma(k)}^TP_{\xi+u,\sigma(k)}\Psi_{\sigma(k)}+\bar{\Phi}_{\sigma(k)}^TQ\bar{\Phi}_{\sigma(k)}
\nonumber
\\  && \quad +\bar{K}_{\sigma(k)}^TR\bar{K}_{\sigma(k)}
\label{eq:Pmmpc}
\end{eqnarray}
where
\begin{eqnarray*}
  \Psi_{\sigma(k)}&=&\tilde{\Phi}_{\sigma(k+m-1)}\cdots\tilde{\Phi}_{\sigma(k+1)}\tilde{\Phi}_{\sigma(k)}\\
  \tilde{\Phi}_{\sigma(k)}&=&\tilde{A}+\tilde{B}_{\sigma(k)}\tilde{K}_{\sigma(k)}\\
  \bar{\Phi}_{\sigma(k)} &=& \left[%
\begin{array}{c}
  \tilde{\Phi}_{\sigma(k)} \\
  \tilde{\Phi}_{\sigma(k+1)}\tilde{\Phi}_{\sigma(k)} \\
  \vdots \\
  \tilde{\Phi}_{\sigma(k+m-1)}\cdots\tilde{\Phi}_{\sigma(k)} \\
\end{array}%
\right] \\
  \bar{K}_{\sigma(k)} &=&\left[%
\begin{array}{c}
  \tilde{K}_{\sigma(k)}\\
  \tilde{K}_{\sigma(k+1)}\tilde{\Phi}_{\sigma(k)} \\
  \vdots \\
  \tilde{K}_{\sigma(k+m-1)}\tilde{\Phi}_{\sigma(k+m-2)}\cdots\tilde{\Phi}_{\sigma(k)} \\
\end{array}%
\right]\\
Q&=&diag(\tilde{q},\ \tilde{q},\cdots, \tilde{q}) \qquad \text{and}
\\ R &=& diag(\tilde{r},\ \tilde{r},\cdots, \tilde{r})
\end{eqnarray*}
}
\end{theorem}

{\bf Proof:}\\
%  The proof detailed below is a straightforward generalization of the time-invariant result.
The closed-loop dynamics of the system \eqref{eq:augmentplant}, when
stabilising linear periodic state feedback of the form
\eqref{eq:mmpclaw} is applied, is given by
\begin{eqnarray*}
  \xi_{k+1} &=& \tilde{A}\xi_{k} +
  \tilde{B}_{\sigma(k)}\Delta\vec{u}_{0|k}
  = (\tilde{A}+\tilde{B}_{\sigma(k)}\tilde{K}_{\sigma(k)})\xi_k
\end{eqnarray*}
or in general
\begin{eqnarray*}
    \xi_{k+i+1} &=& \tilde{\Phi}_{\sigma(k+i)}\xi_{k+i}, \quad
    i=0,1,\ldots
\end{eqnarray*}
where
\begin{eqnarray*}
   \tilde{\Phi}_{\sigma(k+i)} &=&
      \tilde{A}+\tilde{B}_{\sigma(k+i)}\tilde{K}_{\sigma(k+i)}
\end{eqnarray*}
Then
\begin{eqnarray*}
  \left[\begin{array}{c}
  \xi_{k+jm+1}\\ \xi_{k+jm+2}\\ \vdots\\ \xi_{k+jm+m}
  \end{array}\right]
  &=&
  \underbrace{\left[\begin{array}{l}
    \tilde{\Phi}_{\sigma(k)} \\
    \tilde{\Phi}_{\sigma(k+1)}\tilde{\Phi}_{\sigma(k)} \\
    \vdots \\
    \tilde{\Phi}_{\sigma(k+m-1)}\cdots\tilde{\Phi}_{\sigma(k)} \\
  \end{array}\right]}_{\bar{\Phi}_{\sigma(k)}}\xi_{k+jm}
\end{eqnarray*}
and
\begin{eqnarray*}
  \left[\begin{array}{c}
  \Delta\vec{u}_{0|k+jm}\\
  \Delta\vec{u}_{0|k+jm+1}\\
  \vdots\\
  \Delta\vec{u}_{0|k+jm+m-1}
  \end{array}\right]
  &=&
  \left[\begin{array}{l}
  \tilde{K}_{\sigma(k)}\xi_{k+jm}\\
  \tilde{K}_{\sigma(k+1)}\xi_{k+jm+1}\\
  \vdots\\
  \tilde{K}_{\sigma(k+m-1)}\xi_{k+jm+m-1}
  \end{array}\right]\\
  &=&
  \underbrace{\left[\begin{array}{l}
    \tilde{K}_{\sigma(k)}\\
    \tilde{K}_{\sigma(k+1)}\tilde{\Phi}_{\sigma(k)}\\
    \vdots\\
    \tilde{K}_{\sigma(k+m-1)}\tilde{\Phi}_{\sigma(k+m-2)}\ldots\tilde{\Phi}_{\sigma(k)}
  \end{array}\right]}_{\bar{K}_{\sigma(k)}}\xi_{k+jm}
\end{eqnarray*}

Thus,
\begin{eqnarray*}
    J_{\xi,k} &=& \sum_{i=0}^{\infty}\|\xi_{k+i+1}\|^2_{\tilde{q}}\\
    &=& \sum_{j=0}^\infty\sum_{i=1}^m
    \|\xi_{k+jm+i}\|^2_{\tilde{q}}\\
    &=& \sum_{j=0}^\infty \xi_{k+jm}^T\bar{\Phi}_{\sigma(k)}^TQ\bar{\Phi}_{\sigma(k)}\xi_{k+jm} \\
    &=& \xi_k^T [\sum_{j=0}^\infty (\Psi^j_{\sigma(k)})^T
       [\bar{\Phi}_{\sigma(k)}^TQ\bar{\Phi}_{\sigma(k)}]
       \Psi^j_{\sigma(k)} ] \xi_k \\
    &=& \xi_k^T P_{\xi,\sigma(k)} \xi_k
\end{eqnarray*}
where
\begin{eqnarray*}
   Q &=& diag(\tilde{q},\ \tilde{q},\cdots,\tilde{q}) \\
   \Psi_{\sigma(k)}&=&\tilde{\Phi}_{\sigma(k+m-1)}\cdots\tilde{\Phi}_{\sigma(k+1)}\tilde{\Phi}_{\sigma(k)}
\end{eqnarray*}
and
\begin{eqnarray*}
  P_{\xi,\sigma(k)} &=& \sum_{j=0}^\infty (\Psi^j_{\sigma(k)})^T
       [\tilde{\Phi}_{\sigma(k)}^TQ\tilde{\Phi}_{\sigma(k)}]
       \Psi^j_{\sigma(k)}
\end{eqnarray*}
is a convergent series since the controller $\tilde{K}_{\sigma(k)}$
is stabilizing. Thus $P_{\xi,\sigma(k)}$ can be computed by solving
the following Lyapunov equation
\begin{equation*}\label{eq:Pxisigma}
  P_{\xi,\sigma(k)} =
    \Psi_{\sigma(k)}^TP_{\xi,\sigma(k)}\Psi_{\sigma(k)}+\bar{\Phi}_{\sigma(k)}^TQ\bar{\Phi}_{\sigma(k)}
\end{equation*}

Similarly, the sum of the control increments can be computed as
\begin{eqnarray*}
    J_{u,k} &=&
    \sum_{i=0}^{\infty}\| \Delta\vec{u}_{0|k+i}\|^2_{\tilde{r}}
    = \sum_{j=0}^\infty\sum_{i=0}^{m-1} \|
    \Delta\vec{u}_{0|k+jm+i}\|^2_{\tilde{r}}
    = \sum_{j=0}^\infty\sum_{i=0}^{m-1} \| \bar{K}_{\sigma(k+i)}\xi_{k+jm+i}\|^2_{\tilde{r}}\\
    &=& \sum_{j=0}^\infty \xi_{k+jm}^T\bar{K}_{\sigma(k)}^TR\bar{K}_{\sigma(k)}\xi_{k+jm} \\
    &=& \xi_k^T
    \sum_{j=0}^{\infty}(\Psi^j_{\sigma(k)})^T[\bar{K}_{\sigma(k)}^TR\bar{K}_{\sigma(k)}]\Psi^j_{\sigma(k)}\xi_k\\
    &=& x_k^T P_{u,\sigma(k)} x_k
\end{eqnarray*}
where
%\[ J_{u,\sigma(k)} = x_k^T P_{u,\sigma(k)} x_k
%\] where
\begin{eqnarray*}
  R &=& diag(\tilde{r},\ \tilde{r},\cdots, \tilde{r})
\end{eqnarray*}
and
\begin{eqnarray*}
   P_{u,\sigma(k)} &=& \sum_{i=0}^{\infty}(\Psi^i_{\sigma(k)})^T[\bar{K}_{\sigma(k)}^TR\bar{K}_{\sigma(k)}]\Psi^i_{\sigma(k)}
\end{eqnarray*}
which can be computed by solving the Lyapunov equation
\begin{equation*}\label{eq:Pusigma}
   P_{u,\sigma(k)} = \Psi_{\sigma(k)}^TP_{u,\sigma(k)}\Psi_{\sigma(k)}+\bar{K}_{\sigma(k)}^TR\bar{K}_{\sigma(k)}
\end{equation*}
Finally, let
\begin{eqnarray*}
  P_{\xi+u,\sigma(k)} &=& P_{\xi,\sigma(k)} + P_{u,\sigma(k)}\label{eq:Pxiusigma} \\
  &=& \sum_{j=0}^\infty (\Psi^j_{\sigma(k)})^T
       [\bar{\Phi}_{\sigma(k)}^TQ\bar{\Phi}_{\sigma(k)}]
       \Psi^j_{\sigma(k)} +
       \sum_{i=0}^{\infty}(\Psi^i_{\sigma(k)})^T[\bar{K}_{\sigma(k)}^TR\bar{K}_{\sigma(k)}]\Psi^i_{\sigma(k)}
       \nonumber
\end{eqnarray*}
and it is clear that $P_{\xi+u,\sigma(k)}$ can be computed as the
solution of the Lyapunov equation
\[
P_{\xi+u,\sigma(k)}=\Psi_{\sigma(k)}^TP_{\xi+u,\sigma(k)}\Psi_{\sigma(k)}+\bar{\Phi}_{\sigma(k)}^TQ\bar{\Phi}_{\sigma(k)}
\nonumber +\bar{K}_{\sigma(k)}^TR\bar{K}_{\sigma(k)}
\]
\hfill$\blacksquare$
%\end{proposition} %End of Proposition 2

The cost \eqref{eq:lqinfcost}, when the MMPC control law of the form
\eqref{eq:mmpclaw} is applied, is given by \eqref{eq:Jtilde} with
\begin{equation} \label{eq:qrtilde}
\tilde{q}=
\left[\begin{array}{cc}
  q & 0 \\  0 & 0
\end{array}\right], \quad
\tilde{r}= \left[\begin{array}{cc}
  r & 0 \\  0 & 0 \\
\end{array}\right]
\end{equation}
and the corresponding initial conditions hold on $\xi_k$; for
example, at the beginning of an MMPC run it may be appropriate to
set
\[ \xi_k = \left[\begin{array}{cc}
         x_k^T & (\Delta \vec{u}_{k|k-1}^p)^T \\
      \end{array}\right]^T%
   = \left[\begin{array}{cc}
         x_k^T & 0 \\
      \end{array}\right]^T%
\]
%(see Remark~\ref{re:initialstate}).

\begin{remark} \label{rem:4}
Note that the cost defined in \eqref{eq:Jtilde} differs in the first
term from that defined in \eqref{eq:mmpc2a}. That is, in
\eqref{eq:Jtilde} we do not include any contribution from $x_{k|k}$,
since that is fixed and cannot be influenced by the optimisation at
time step $k$.
\end{remark}

\begin{remark}
It is seen that the system \eqref{eq:augmentplant} is linear and
periodic, while the cost \eqref{eq:Jtilde} is quadratic with
constant coefficients. Thus the optimal control law can be obtained
from the theory given in \cite{BitColDen:88}, and is of the form
\eqref{eq:mmpclaw}. This suggests yet another method to compute the
terminal cost $F_{\sigma(k)}(x_{k+N|k})$ to ensure nominal stability
of MMPC in addition to that presented in Theorem
\ref{thm:nominalMMPC}. A family of MMPC designs may be obtained by
optimising the cost function \eqref{eq:Jtilde} subject to the system
\eqref{eq:augmentplant} by choosing appropriate $\tilde{q}$ and
$\tilde{r}$ matrices.
\end{remark}

\begin{remark}
If the second part of $\xi_k$, namely $\Delta \vec{u}_{k|k-1}^p$,
were included in the optimisation, so that the optimal cost became a
function of $x_k$ only, then the optimal cost, and the optimal
solution, would be the same as that obtained with `scheme 1' in our
earlier papers, namely it would correspond to the cost resulting
from allowing each `agent' to optimise \emph{all} future inputs
rather than just `its own' input.
\end{remark}

\begin{ignore}
\begin{remark}
To obtain the optimal unconstrained MMPC cost, the state feedback
gains $\tilde{K}_{\sigma(k+i)}$ $(i=0,\ldots,m-1)$ are obtained by
applying \eqref{eq:periodicRDE} and \eqref{eq:LQK} with parameters
taken from \eqref{eq:augmentplant}, \eqref{eq:Jtilde} and
\eqref{eq:qrtilde}.
\end{remark}

\begin{remark}
The analysis in this section suggests that for {\it unconstrained}
MMPC, standard periodic control framework can be used to compute the
MMPC gains via the augmented state space model
(Eq.\ref{eq:augmentplant}). More specifically, the unconstrained
MMPC gain is given by { \small
\begin{eqnarray*}
  \tilde{K}_{\sigma(k)} &=&
     -(\tilde{B}_{\sigma(k)}^T\tilde{P}_{\sigma(k+1)}\tilde{B}_{\sigma(k)}+\tilde{r})^{-1}
     \tilde{B}_{\sigma(k)}^T\tilde{P}_{\sigma(k+1)}\tilde{A}
\end{eqnarray*}
}

where $\tilde{P}_{\sigma(k+1)}$ is the SPPS solution of the
discrete-time periodic Riccati equation {\small
\begin{eqnarray*}
    \tilde{P}_k &=& \tilde{A}^T \tilde{P}_{k+1}\tilde{A}
    - \tilde{A}^T \tilde{P}_{k+1}\tilde{B}_{\sigma(k)}(\tilde{B}_{\sigma(k)}^T
    \tilde{P}_{k+1}\tilde{B}_{\sigma(k)}+\tilde{r})^{-1} \\
      && \quad \times B_{\sigma(k)}^T \tilde{P}_{k+1}\tilde{A}+\tilde{q}
\end{eqnarray*}
}

%When the active constraints are known, the cost of MMPC can be
%computed by solving the corresponding discrete-time periodic Riccati
%equation, \small
%\begin{equation*}
%    \tilde{P}_k = \tilde{A}^T \tilde{P}_{k+1}\tilde{A} - \tilde{A}^T \tilde{P}_{k+1}\tilde{B}_{\sigma(k)}(\tilde{B}_{\sigma(k)}^T \tilde{P}_{k+1}\tilde{B}_{\sigma(k)}+\tilde{r})^{-1}
%            B_{\sigma(k)}^T \tilde{P}_{k+1}\tilde{A}+\tilde{q}
%\end{equation*}
%\normalsize with the optimal periodic controller gains
%$\tilde{K}_{\sigma(k)}$ being
%\[
%\tilde{K}_{\sigma(k)}=-(\tilde{B}_{\sigma(k)}^T\tilde{P}_{\sigma(k+1)}\tilde{B}_{\sigma(k)}+\tilde{r})^{-1}\tilde{B}_{\sigma(k)}^T\tilde{P}_{\sigma(k+1)}\tilde{A}
%\]
\end{remark}
\end{ignore}

%%% This is file "robustmmpc.tex".

%%% Edited by KVL, 16.02.10.
%%% Edited by JMM, 20.12.09.
%%% Edited by JMM, 22.08.09.
%%% Edited by JMM 29.07.09 (minor corretions)
%%% Edited from ECC'07 paper ("ecc07_mmpc.tex") by JMM, 18.12.2007.
%%% ECC'07 paper was originally written mostly by AGR.
%%% Edited by KVL, 24 April 2007 in Cambridge
%%% Edited by JMM, 18 July 2008 at NTU.
%%% Edited by AGR, 1 Aug 2008, in Bristol
% Edited by JMM, 18.09.08, 13.10.08.

\section{Robust MMPC} \label{robustmmpc}
This section  develops a robust version of MMPC. Uncertainty is
introduced into the plant model as a bounded disturbance. The
constraints which appear in the MMPC algorithm are then modified so
that robust feasibility can be guaranteed, providing that it is
achieved initially.
%Stability can be established using the results of
%Refs.~\cite{LMW05,LMW06} to evaluate the cost function for the nominal
%case, together with those of Ref.~\cite{RH06a} to take account of
%constraint tightening. We omit the details for the sake of brevity.

%The main contribution of this paper is Scheme~2,
%The main interest of this section is Scheme~2, which offers both
%performance and computational benefits over conventional MPC.
%However, Scheme~1 is still required to initialize the Scheme~2
%controller.

%\subsection{Problem Statement} \label{sec:probstat}
%
The plant dynamics \eqref{sysA} are now extended to include an
unmeasured but bounded disturbance $w_k$:
\begin{equation}
x_{k+1} = Ax_{k}+\sum_{j=1}^m B_j \Delta u_{j,k} + Ew_{k}.
\label{eq:dyn1}
\end{equation}
where $w_{k}$ satisifies
\begin{equation}
w_k \in \mathcal{W} \ \forall k
\label{eq:wbound}
\end{equation}
and $\mathcal{W}$~is a known, bounded set containing the origin.
%The system is required to
%obey the following state constraints
%%
%\begin{equation}
%%
%x_k \in \mathcal{X} \ \forall k.
%\label{eq:consreq}
%%
%\end{equation}

As explained in Section~\ref{formulation}, the system dynamics
\eqref{eq:dyn1} can be re-written as a periodic linear system
\begin{equation}
x_{k+1} = Ax_{k}+B_{\sigma(k)} \Delta \tilde u_k + Ew_{k}
\label{eq:dyn2}
\end{equation}

\begin{ignore} %KVL:24April08
The following indexing function identifies the input channel to
be moved at each step
\begin{equation}
\sigma(k) = (k \bmod m)+1
\end{equation}
so that the asynchronous nature of the multiplexed control moves, as
illustrated in Fig.~\ref{fig:inputpattern}, is captured by the
constraint
\begin{equation}
\Delta u_{j,k} = 0 \ \mathrm{if} \ j \neq \sigma(k).
\end{equation}
It is then possible to rewrite the system dynamics~(\ref{eq:dyn1}) as
a periodic time-varying system
\begin{equation}
x_{k+1} = Ax_{k}+B_{\sigma(k)} \Delta \tilde u_k + Ew_{k}
\label{eq:dyn2}
\end{equation}
where~$\Delta \tilde u_k = \Delta u_{\sigma(k),k}$. This form will
be used for predictions in the optimizations and illustrates how
MMPC can draw on results for periodic time-varying systems.
\end{ignore}

%\subsection{Controller Formulations} \label{sec:controllers}
%

%\subsubsection{Scheme 1} \label{sec:scheme1}

%\fbox{OMIT THIS? WE HAVE OMITTED SCHEME 1 IN PREVIOUS SECTION.}

For robust MMPC we solve the following finite-time constrained
linear periodic control problem, % $\mathcal{P}_{\sigma(k)}(x_k)$:
which we denote
$\mathcal{P}_{\sigma(k)}(x_{k|k},\Delta\vec{u}^p_{k|k-1},w_{k-1})$:
{\small
\begin{equation} \label{eq:robustmmpc}
  \begin{array}{ll}
    \mathrm{Minimize}&
  J_k = F_{\sigma(k)}(x_{k+N|k})+
     \sum_{i=0}^{N-1} \left(\| x_{k+i|k} \|_q^2 +
       \| \tilde{u}_{k+i|k} \|_r^2\right)\\
  \mathrm{wrt}& \Delta \tilde{u}_{k+i|k} \quad(i=0,m,2m,\ldots,N-1) \\
  \mathrm{s.t.} & \Delta \tilde{u}_{k+i|k}\in\mathcal{U}_{i,\sigma(k)}, \quad(i=0,1,\ldots,
  N-1)\\
  & x_{k+i|k}\in\mathcal{X}_{i,\sigma(k)}, \quad(i=1,2,\ldots,N-1) \\ %\mathbb{X}, \forall i\in\{1,\ldots, N\} \\
  & x_{k+N|k}\in\mathcal{T}_{\sigma(k)} \\ %\mathcal{X}_I(K_{\sigma(k)}) \\
%  & -K_{\sigma(k+1)}x_{k+N|k} \in\mathcal{U}_{N,\sigma(k)} \\
  & x_{k+i+1|k} = Ax_{k+i|k} + {B}_{\sigma(k+i)}\Delta\tilde{u}_{k+i|k}, \quad(i=0,1,\ldots) \\
  & \Delta \tilde u_{k+i|k} = \Delta \tilde u_{k+i|k-1} + M_{i,\sigma(k)}Ew_{k-1}, \ \forall i \neq jm
  \end{array}
\end{equation}
}

Note some differences from \eqref{eq:mmpc2a}. The predicted inputs
and states are constrained to lie in sets
$\mathcal{U}_{i,\sigma(k)}$ and $\mathcal{X}_{i,\sigma(k)}$ which
depend on how far into the prediction horizon they are, as well as
on $k$. The target set at the end of the horizon has been modified
from $\mathcal{X}_{\cal{I}}(K_{\sigma(k)})$ to
$\mathcal{T}_{\sigma(k)}$. Finally, the inputs which are not being
optimised are assumed to be modified from their previously planned
values by the feedback term $M_{i,\sigma(k)}Ew_{k-1}$; note that the
value of $Ew_{k-1}$ can be inferred from data $\{u_{j-1},x_j: j\leq
k\}$.

%$F(x_{k+N})$ is a suitably chosen terminal cost, and $\mathbb{X}$
%and $\mathbb{U}$ are compact polyhedral sets containing the origin
%in their interior. $\mathcal{X}_I(K_{\sigma(k)})$ denotes the sets
%in which none of the constraints is active, and which is the maximum
%positively invariant set (see, for example,~\cite{Bla:99}) for the
%linear periodic system (\ref{eq:periodicplant}), when a stabilizing
%linear periodic feedback controller $K_{\sigma(k)}$ is applied,
%namely
%\begin{eqnarray*}
%  && x_k\in\mathcal{X}_I(K_{\sigma(k)})
%  \Rightarrow K_{\sigma(k)}x_k \in \mathbb{U}\ \mathrm{and} \\
%  && (A-B_{\sigma(k)}K_{\sigma(k)})x_k\in \mathcal{X}_I(K_{\sigma(k)})
%\end{eqnarray*}
%where $\mathcal{X}_I(K_{\sigma(k)})\subset \mathbb{X}$.

\begin{ignore} %KVL, 24Apr08
The controller for Scheme~1 is based on an optimisation of the control
moves of all channels over a horizon of $N$~steps. The decision
variable is therefore~$\mathbf{\Delta U}_k=(\Delta \tilde u_{k|k} \
\Delta \tilde u_{k+1|k} \ \Delta \tilde u_{k+2|k} \ \ldots \ \Delta
\tilde u_{k+N-1|k} )^T$ where~$\Delta \tilde u_{k+i|k}$ denotes the
prediction made at time~$k$ of a control move to be executed at
time~$k+i$. The optimization solved at every step is
\begin{equation}
\min_{\mathbf{\Delta U}_k} V_{\sigma(k+N)}(x_{k+N|k}) +
\sum_{i=0}^{N-1} || x_{k+i|k} ||_q^2 + || \Delta \tilde u_{k+i|k}
||_r^2
\label{eq:sch1opt}
\end{equation}
subject to~$\forall i \in \{ 0, \ldots, N-1 \}$
\begin{subequations}
\begin{equation}
x_{k+i+1|k} = Ax_{k+i|k} + B_{\sigma(k+i)}\Delta \tilde u_{k+i|k}
\end{equation}
\begin{equation}
x_{k+N|k} \in \mathcal{T}_{\sigma(k)}
\end{equation}
\begin{equation}
x_{k|k} = x(k)
\end{equation}
\begin{equation}
x_{k+i|k} \in \mathcal{X}_{i,\sigma(k)}
\end{equation}
\label{eq:cons}%
\end{subequations}
The notations~$\| \cdot \|_q$ and~$\| \cdot \|_r$ represent typical
weighted quadratic costs.  The terminal
penalty~$V_{\sigma(k+N)}(x_{k+N|k})$ represents the cost to complete
the problem from the predicted state at the horizon~$x_{k+N|k}$ and a
suitable cost-to-go function was derived in Ref.~\cite{LMW05}.
\end{ignore}

The constraint
sets~$\mathcal{U}_{i,\sigma(k)}$,\,$\mathcal{X}_{i,\sigma(k)}$
and~$\mathcal{T}_{\sigma(k)}$ are constructed to ensure robust
feasibility, such that if some solution
\begin{equation} \label{eq:Ustar}
\Delta\vec{U}^*_{k|k} = \left( \Delta \tilde u_{k|k}^{*} , \ \Delta
\tilde u_{k+1|k}^{*} , \ \Delta\tilde{u}_{k+2|k}^*, \ \ldots, \
\Delta \tilde u_{k+N-1|k}^{*} \right)^T
\end{equation}
is feasible at some time $k$ then a candidate solution
\begin{equation}
\widehat{\Delta U}_{k+1} = \left(
\begin{array}{@{}r@{}c@{}l@{}}
\Delta \tilde u^*_{k+1|k} &+& M_{0,\sigma(k+1)}Ew_k\\
&\vdots& \\
\Delta \tilde u^*_{k+N-1|k} &+& M_{N-2,\sigma(k+1)}Ew_k\\
-K_{\sigma(k+1)} x^*_{k+N|k} &+& M_{N-1,\sigma(k+1)}Ew_k
\end{array}
\right)
\label{eq:cand}%
\end{equation}
is feasible at time~$k+1$ for all~$w_k \in \mathcal{W}$. The designer
chooses the feedback parameters~$M_{i,\sigma(k)}$ and~$K_{\sigma(k)}$
offline (as in~\cite{RH06a}, on which this development is based).

To achieve this robust feasibility property, the state constraints
$x_k\in \mathbb{X}$ are tightened using a recursion
\begin{subequations}
\begin{equation} \label{eq:XX0}
\mathcal{X}_{0,\sigma(k)} = \mathbb{X}
\end{equation}
\begin{equation} \label{eq:XXW}
\mathcal{X}_{i+1,\sigma(k)} = \mathcal{X}_{i,\sigma(k+1)} \sim
L_{i,\sigma(k+1)}E\mathcal{W}
\end{equation}
\label{eq:ct}%
\end{subequations}
where
\begin{subequations}
\begin{equation} \label{eq:L0k}
L_{0,\sigma(k)} = I
\end{equation}
\begin{equation} \label{eq:LALBF}
L_{i+1,\sigma(k)} = A L_{i,\sigma(k)} + B_{\sigma(k+i)} M_{i,\sigma(k)}
\end{equation}
\label{eq:ldef}%
\end{subequations}
for the chosen feedback policy $M_{i,\sigma(k)}$ and the
``$\sim$''~operator denotes the Pontryagin difference %~\cite{KG98}
between two sets:
\begin{equation}
\mathcal{A} \sim \mathcal{B} = \{ a \ | \ a+b \in \mathcal{A} \
\forall b \in \mathcal{B} \}
\label{eq:pontrydef}
\end{equation}
Similarly, the input move constraint sets $\Delta \tilde{u}_k \in
\mathbb{U}_k$ are tightened using the following recursion
\begin{subequations}
\begin{equation} \label{eq:robUrecursion0}
    \mathcal{U}_{0,\sigma(k)} = \mathbb{U}_{\sigma(k)}
\end{equation}
\begin{equation}\label{eq:robUrecursion}
    \mathcal{U}_{i,\sigma(k)} = \mathcal{U}_{i-1,\sigma(k+1)} \sim
    M_{i-1,\sigma(k+1)}E\mathcal{W}
\end{equation}
\end{subequations}
The terminal sets~$\mathcal{T}_{\sigma(k)}$ have the robust
invariance properties that, if $x \in \mathcal{T}_{\sigma(k)}$ and
$w \in \mathcal{W}$ then
\begin{subequations}
\begin{equation} \label{eq:Tinvariance1}
\left(A - B_{\sigma(k+N)} K_{\sigma(k+1)}\right) x +
\left[AL_{N-1,\sigma(k+1)} + B_{\sigma(k+N)}M_{N-1,\sigma(k+1)}
\right] Ew \in \mathcal{T}_{\sigma(k+1)}
\end{equation}
\begin{equation}
  -K_{\sigma(k+1)}x \in\mathcal{U}_{N,\sigma(k)}
\end{equation} and
\begin{equation} \label{eq:Tinvariance2}
\mathcal{T}_{\sigma(k)} \subseteq \mathcal{X}_{N,\sigma(k)}.
\end{equation}
\end{subequations}
%
%The reader is directed to Ref.~\cite{RH06a} for a more thorough
%explanation of how these constraint modifications imply feasibility of
%the solution in~(\ref{eq:cand}).

%Note that the union of these $m$~terminal sets is invariant over
%$m$~steps, since $\sigma(k+m)=\sigma(k)$: this property is the
%analogue of the familiar invariant terminal set requirement for the
%stability of predictive control~\cite{MRRS00}.

The parameters~$M_{i,\sigma(k)}$ and~$K_{\sigma(k)}$ are chosen by
the designer. The parameters~$L_{i,\sigma(k)}$, which relate the
control perturbations in~(\ref{eq:cand}) to the corresponding
changes in the state predictions, are then fixed by~(\ref{eq:ldef}).
These settings determine the amount of constraint tightening applied
in~(\ref{eq:ct}). Typically, to achieve a large feasible region, the
control policy chosen should minimise the quantities limited by the
constraints.

A restrictive but convenient choice of candidate policy is to
select~$M_{i,\sigma(k)}, \ i=0,\ldots,N-2$ such that~$L_{N,\sigma(k)} =
0 \ \forall k$ and then set~$M_{N-1,\sigma(k)}=0$, $K_{\sigma(k)}=0$
and $\mathcal{T}_{\sigma(k)} = \{0\} \ \forall k$.

The following algorithm defines robust MMPC. It uses notations
defined in \eqref{sysA}, \eqref{eq:periodicplant} and
\eqref{eq:delvecuk0k}. It is the same as Algorithm \ref{alg:MMPC}
except that problem \eqref{eq:robustmmpc} is solved instead of
problem \eqref{eq:mmpc2a}.

\begin{algorithm}[Robust MMPC] \label{alg:robMMPC}
\begin{enumerate}
\item[]
\item \label{step:robloop1} Set $k := k_0$. Initialise by solving
problem~\eqref{eq:robustmmpc}, but optimising over all the variables
$\Delta\tilde{u}_{k+i|k}, i=0,1,\ldots,N-1$.
\item \label{step:robloop2} Apply control move $\Delta u_{\sigma(k),k} = \Delta \tilde u_{k|k}$
\item Store planned moves $\Delta\vec{u}_{k,m|k}$.
\item Pause for one time step, increment $k$, obtain new measurement
$x_k$.
\item Solve problem~\eqref{eq:robustmmpc}.
\item  Go to step~\ref{step:robloop2}.
\end{enumerate}
 \end{algorithm}
%
%Note that Step~\ref{step:robloop1} involves solving for inputs
%across all channels, not just channel~$\sigma(k)$.  This type of
%initialisation requirement is common in distributed MPC.  Subsequent
%results do not depend on the optimality of this initial solution,
%only its feasibility.

We will need the following result concerning the use of the
$L_{i,\sigma(k)}$ matrices.
\begin{lemma} \label{thm:robMMPClemma}
Suppose that $x_{k+1} = x_{k+1|k} + E w_k$ and
\begin{equation}\label{eq:lemmaAssumtion}
 \Delta\tilde{u}_{k+j|k+1} = \Delta\tilde{u}_{k+j|k} +
 M_{j-1,\sigma(k+1)}Ew_k, \quad (j=1,2,\ldots,i)
\end{equation}
Then
\begin{equation} \label{eq:lemma}
 x_{k+i|k+1} = x_{k+i|k}+L_{i-1,\sigma(k+1)}Ew_k, \quad
 (i=1,2,\ldots)
\end{equation}
\end{lemma}

\textbf{Proof:} We prove the lemma by induction on $i$. Suppose the
result is true for some $i$. Then
\begin{eqnarray}
 x_{k+i+1|k+1} &=& Ax_{k+i|k+1} + B_{\sigma(k+i)}
 \Delta\tilde{u}_{k+i|k+1} \\
 &=& A[x_{k+i|k}+L_{i-1,\sigma(k+1)}Ew_k] + B_{\sigma(k+i)}\Delta\tilde{u}_{k+i|k+1}
\end{eqnarray}
But
\begin{equation}\label{}
 x_{k+i+1|k} = Ax_{k+i|k} + B_{\sigma(k+i)}\Delta\tilde{u}_{k+i|k}
\end{equation}
and, by assumption,
\begin{equation}\label{}
 \Delta\tilde{u}_{k+i|k+1} = \Delta\tilde{u}_{k+i|k} + M_{i-1,\sigma(k+1)}Ew_k
\end{equation}
so that
\begin{eqnarray}
 x_{k+i+1|k+1} &=& x_{k+i+1|k}+ [AL_{i-1,\sigma(k+1)} +
 B_{\sigma(k+i)}M_{i-1,\sigma(k+1)}]Ew_k \\
 &=& x_{k+i+1|k}+ L_{i,\sigma(k+1)}Ew_k \quad\text{because of \eqref{eq:LALBF}}
\end{eqnarray}
and hence the claimed result is true for $i+1$.

Now consider $i=1$: $x_{k+1|k+1} = x_{k+1} = x_{k+1|k}+Ew_k$, so the
claimed result holds for $i=1$, since $L_{0,\sigma(k+1)}=I$, by
definition \eqref{eq:L0k}.

Thus the result holds for $i\geq 1$. \hfill $\blacksquare$

\begin{theorem}
If the system~(\ref{eq:dyn1}) is controlled using
Algorithm~\ref{alg:robMMPC} and the initial optimisation at
time~$k=k_0$ (\emph{ie} step 1 of the algorithm) can be solved, and
$x_{k_0}\in \mathbb{X}$, then $(i)$~the optimisation remains feasible
and $(ii)$~the constraints $x_k \in \mathbb{X}$ and $\Delta \tilde{u}_k
\in \mathbb{U}_k$ are satisfied for $k>k_0$ and for all disturbances
satisfying~(\ref{eq:wbound}).
\end{theorem}

\textbf{Proof:} $(i)$~We will begin by showing that, by construction of the
constraints in~(\ref{eq:ct}), feasibility at any time~$k$ implies
feasibility at time~$k+1$. In particular, we will demonstrate
feasibility by establishing that the candidate
solution~(\ref{eq:cand}) satisfies all the constraints of the
optimisation. Therefore, feasibility at time~$k=k_0$ implies
feasibility at all future times~$k>k_0$.

Assume that we have a feasible solution \eqref{eq:Ustar} at time
$k$, %that $x_k\in \mathbb{X}$,
and that $\Delta\tilde{u}^*_{k|k}$ is
applied as input to the plant \eqref{eq:dyn2}.  This results in the
next plant state being
\begin{equation}
 x_{k+1} = Ax_k +
 B_{\sigma(k)}\Delta\tilde{u}^*_{k|k} + Ew_k = x_{k+1|k}+Ew_k
\end{equation}
Thus from~(\ref{eq:cand}) and Lemma~\ref{thm:robMMPClemma} we have
\begin{subequations}
\begin{eqnarray}
x_{k+i+1|k+1} &=& x_{k+i+1|k}+ L_{i,\sigma(k+1)}Ew_k\\
\Delta\tilde{u}_{k+i+1|k+1} &=& \Delta\tilde{u}_{k+i+1|k} +
 M_{i,\sigma(k+1)}Ew_k  \ (i=0,1,\ldots,N-1)
\end{eqnarray}
\label{eq:perturbed}%
\end{subequations}
and since \eqref{eq:Ustar} was assumed feasible, we know
$x_{k+i+1|k} \in \mathcal{X}_{i+1,\sigma(k)}$ and
$\Delta\tilde{u}_{k+i+1|k} \in \mathcal{U}_{i+1,\sigma(k)}$.
Combining this with~\eqref{eq:perturbed}, the definition of the
Pontryagin difference~\eqref{eq:pontrydef} and the
recursions~\eqref{eq:XXW} and~\eqref{eq:robUrecursion}, we
know~$x_{k+i+1|k+1} \in \mathcal{X}_{i,\sigma(k+1)}$
and~$\Delta\tilde{u}_{k+i+1|k+1} \in \mathcal{U}_{i,\sigma(k+1)}$
for all $w_k \in \mathcal{W}$.

We also need to show that $x_{k+N+1|k+1}\in
\mathcal{T}_{\sigma(k+1)}$ if the candidate solution \eqref{eq:cand}
is applied. We have $x_{k+N|k}\in \mathcal{T}_{\sigma(k)}$ by
the assumption of feasibility at time~$k$.
\begin{equation}\label{eq:xkN1k1}
    x_{k+N+1|k+1} =
    Ax_{k+N|k+1}+B_{\sigma(k+N)}\Delta\tilde{u}_{k+N|k+1}
\end{equation}
But, from \eqref{eq:cand},
\begin{equation}\label{eq:ukNk1}
    \Delta\tilde{u}_{k+N|k+1} =
    -K_{\sigma(k+1)}x_{k+N|k}+M_{N-1,\sigma(k+1)}Ew_k
\end{equation}
and, from Lemma \ref{thm:robMMPClemma} (since
$\Delta\tilde{u}_{k+N|k}=-K_{\sigma(k+1)}x_{k+N|k}$),
\begin{equation}\label{eq:xkNk1}
    x_{k+N|k+1} = x_{k+N|k} + L_{N-1,\sigma(k+1)}Ew_k
\end{equation}
Hence, substituting \eqref{eq:ukNk1} and \eqref{eq:xkNk1} into
\eqref{eq:xkN1k1} gives
\begin{eqnarray}
    x_{k+N+1|k+1} &=&
    \left[A-B_{\sigma(k+N)}K_{\sigma(k+1)}\right]x_{k+N|k} +
    \nonumber \\
    && +
    \left[AL_{N-1,\sigma(k+1)}+B_{\sigma(k+N)}M_{N-1,\sigma(k+1)}\right]Ew_k
    \\
    &\in& \mathcal{T}_{\sigma(k+1)} \quad\text{because of \eqref{eq:Tinvariance1}.}
\end{eqnarray}
Having established~$x_{k+i+1|k} \in \mathcal{X}_{i+1,\sigma(k)}$,
$\Delta\tilde{u}_{k+i+1|k} \in \mathcal{U}_{i+1,\sigma(k)}$
and~$x_{k+N+1|k+1} \in \mathcal{T}_{\sigma(k+1)}$ for all $w_k \in
\mathcal{W}$, the feasibility of the candidate solution has been
proven, and thus the feasibility of the optimisation is proven.

$(ii)$~It remains to show that the state and input constraints are
satisfied.  Feasibility at all steps demands that~$x_k = x_{k|k} \in
\mathcal{X}_{0,\sigma(k)}$ which from~\eqref{eq:XX0} implies $x_k \in
\mathbb{X}$.  Similarly, $\Delta \tilde{u}_k = \Delta\tilde{u}_{k|k}
\in \mathcal{U}_{0,\sigma(k)}$ which from~\eqref{eq:robUrecursion0}
implies $\Delta \tilde{u}_k \in \mathbb{U}_k$.

 \hfill $\blacksquare$

\section{Examples} \label{example}

This section demonstrates the potential benefits of MMPC by
employing it in simulation for the control of three different
example systems. In the first example we consider nominal MMPC; we
show how the cost formula can be used to evaluate some of the design
choices.
 In the second and third examples comparisons are made between the robust MMPC
scheme and standard --- but also robustified --- ``synchronous''
MPC~(SMPC). In these examples all simulations were performed on the
same PC with a 3.2GHz Intel Pentium~4 processor and 1GB RAM. Matlab
version~7.1 (R14, Service Pack~3) was employed, using Simulink to
simulate the system dynamics and the ``quadprog'' optimisation
function to solve the necessary quadratic programming~(QP) problems.
Computation times were measured using the Matlab profiler.

\subsection{Nominal MMPC: Effects of $\mathbf{N_u}$ and updating sequence}%\label{sec:example}
 In this section,
numerical examples will be given to illustrate how the cost formula
for MMPC can be used for evaluating the effect of various values of
$N_u$, and of the updating sequence, on the closed-loop performance,
when constraints are not active.

The cost formula for MMPC is calculated as (\ref{eq:Jtilde}) with
(\ref{eq:qrtilde}) and initial condition of
$\xi_k=\left[\begin{array}{cc}
         x_k^T & 0 \\
      \end{array}\right]^T$.
Hence, only the upper-left $n\times n$ ($n$ is the dimension of
$x_k$) sub-matrices are relevant in the cost computation. To be
specific, the sub-matrices are $\hat{P}_{\sigma(k)}$,
$\hat{P}_{x,\sigma(k)}$ and $\hat{P}_{u,\sigma(k)}$ as shown below
\[
P_{\xi+u,\sigma(k)}=\left[%
\begin{array}{cc}
  \hat{P}_{\sigma(k)} & \star\\
 \star & \star \\
\end{array}%
\right],\quad P_{\xi,\sigma(k)}= \left[%
\begin{array}{cc}
  \hat{P}_{x,\sigma(k)} & \star \\
 \star & \star \\
\end{array}%
\right],\]

\[
P_{u,\sigma(k)}=\left[%
\begin{array}{cc}
  \hat{P}_{u,\sigma(k)} & \star \\
 \star & \star \\
\end{array}%
\right]
\]
where $P_{\xi+u,\sigma(k)}$, $P_{\xi,\sigma(k)}$ and
$P_{u,\sigma(k)}$ are defined in \eqref{eq:Pmmpc}, \eqref{eq:Pxi}
and \eqref{eq:Pu}, respectively, and $\star$ denotes a sub-matrix of
compatible dimensions, which can be omitted from the cost
computation.

Therefore, the quadratic cost of MMPC can be computed as
\[
  J_{\sigma(k)} =  J_{x,\sigma(k)}+J_{u,\sigma(k)}=x_k^T\hat{P}_{\sigma(k)}x_k
\]
where $J_{x,\sigma(k)} = x_k^T\hat{P}_{x,\sigma(k)}x_k$ and
$J_{u,\sigma(k)} = x_k^T\hat{P}_{u,\sigma(k)}x_k$.

%where $P_{\sigma(k)}$, $P_{x,\sigma(k)}$ and $P_{u,\sigma(k)}$ are
%extract from $P_{\xi+u,\sigma(k)}$, $P_{\xi,\sigma(k)}$ and
%$P_{u,\sigma(k)}$ accordingly with proper dimension.

Now we have a way to compare different MMPC schemes, including
differences in horizon lengths and update sequences. In other words,
given two MMPC schemes, whose costs are $J_i=x_k^T\hat{P}_ix_k$ and
$J_j=x_k^T\hat{P}_jx_k$, then
\[
J_i-J_j=x_k^T(\hat{P}_i-\hat{P}_j)x_k
\]
Hence analysis of the properties of the difference
$\hat{P}_i-\hat{P}_j$ gives information on the relative merits of
the two MMPC designs. For example, $\hat{P}_i-\hat{P}_j>0$ indicates
that design $j$ is better than design $i$ for all initial conditions $x_0$.
%\fbox{IS $trace(P_i-P_j)$ USEFUL?}
%So, from the eigenvalues of $(\hat{P}_i-\hat{P}_j)$, we could
%compare the two MMPC schemes.

Consider the following two-input-two-output continuous-time plant
\begin{equation*}
  \left[\begin{array}{c}
    y_1(s) \\
    y_2(s) \\
  \end{array}\right] = \left[\begin{array}{cc}
                                \frac{1}{7s+1} & \frac{1}{3s+1}\\
                                \frac{2}{8s+1} & \frac{1}{4s+1} \\
                             \end{array}\right]
                             \left[\begin{array}{c}
                                    u_1(s) \\
                                    u_2(s) \\
                                   \end{array}\right]
\end{equation*}

We chose the sampling time to be $T = 1s$. For MMPC, the states were
measured at $T/m=0.5s$ with $u_1$ and $u_2$ alternatively applied at
$0.5s$ intervals, but each held constant over a period of $T=1s$,
that is, $u_1$ is updated at times $(0, 1s,2s,\ldots)$ and $u_2$ is
updated at times $(0.5s, 1.5s, \ldots)$ (for the updating sequence
of $u_1,\ u_2, \ u_1,\ u_2, \cdots$,). For all the results listed
below, the tuning parameters for MMPC are: $q=I$ and $r=1$.

%\begin{itemize}

%\item \textbf{The effect of tuning parameter $N_u$ in MMPC.}
%\subsection{The effect of tuning parameter $N_u$ in MMPC}
The horizon length $N_u$ is an important tuning parameter for MPC in
general, and remains so for MMPC. With the MMPC cost formula, we can
compute the cost and predict the performance difference with
different $N_u$. To illustrate, we generate a simulation scenario by
adding a step input disturbance to the plant and see how the
performance and cost vary with different $N_u$. This is done by
replacing $\Delta u$ in \eqref{sysA} by $(\Delta u + \Delta d)$,
namely modelling a \emph{change} of input disturbance, so that an
impulse on $\Delta d$ corresponds to a step disturbance. This then
allows such a disturbance to be represented by the initial condition
$x(0)=B\Delta d$, which in turn allows the use of formula
\eqref{eq:Jtildesum}. (The results were checked against numerical
estimation of the cost accumulated during simulation.)

MMPC is a periodic control scheme; thus its performance depends on
the time at which a disturbance occurs. More specifically, for the
two-input plant considered here, depending on the time at which
disturbances occur, $u_{1,k}$ may react first (ie control updating
sequence $(u_{1,k}, u_{2,k}, u_{1,k+1} \ldots)$), or $u_{2,k}$ may
react first (ie control updating sequence $(u_{2,k}, u_{1,k},
u_{2,k+1}, \ldots)$). This depends on whether $\sigma(k)=1$ or
$\sigma(k)=2$. This section uses the cost formula
\eqref{eq:Jtildesum} to compare the cost of MMPC for these two
different updating sequences in a specific scenario.

%MMPC is a periodic control scheme with flexible choice of updating
%patterns. This section will show how the updating sequence affects
%the cost and the
%closed-loop performance. %with the same system setting. %We will still
%use the step input disturbance as a specific simulation scenario.

Table \ref{tab:DifNuEigSeq} shows the eigenvalues of
$\hat{P}_1-\hat{P}_2$ as $N_u$ varies, where $\hat{P}_1$ represents
the cost matrix when $\sigma(k)=1$  (updating sequence
$(u_{1,k},u_{2,k},\ldots)$) while $\hat{P}_2$ represents the cost
matrix when $\sigma(k)=2$ (updating sequence
$(u_{2,k},u_{1,k},\ldots)$). From the table, it can be seen that
$\hat{P}_1-\hat{P}_2$ is indefinite, which means that one updating
sequence is not definitely better than the other, but depends on the
specific scenario --- as expected. Fig. \ref{fi:DifSeqNu5} compares
the closed-loop performance between the two updating sequences when
$N_u=5$. The solid lines show the response to a step disturbance on
each input when $u_1$ is the first input to react to it (disturbance
occurs at step $k$ and $\sigma(k)=1$), while the dashed lines show
the response when $u_2$ is the first input to react to the
disturbance. The input trajectories approximately interchange in the
two cases, as do the output trajectories, so there is little to
choose between the two as regards performance.
%It can be seen that in the first case the overshoot of
%$y_1$ is greater than in the second case, while the undershoot is
%smaller in the first case. For output channel $y_2$, the first case
%shows a bigger undershoot and a smaller overshoot. It is hard to
%decide which one is better simply from the simulation plots.
%The eigenvalues of $(\hat{P}_1-\hat{P}_2)$ are $[ -28.3386,\
%-0.0710,\ -0.0001,\ 0.0000,\ 0.0905,\ 18.2002]$.
The cost difference
$x_k^T(\hat{P}_1-\hat{P}_2)x_k$ in this case is $0.3599$, which
means that the second updating sequence is slightly better than the
first for this particular disturbance, as judged by the cost
function.

\begin{table}[htp]
\caption{MMPC: Eigenvalues of $(\hat{P}_1-\hat{P}_2)$ with different
$N_u$}\label{tab:DifNuEigSeq}
\begin{center} {
\begin{tabular}{|c|cccccc|}
  \hline
  $N_u$ & \multicolumn{6}{c|}{Eigenvalues of $\displaystyle (\hat{P}_1-\hat{P}_2)$} \\
 \hline
  $1$  &  -9.4274   &  -0.0069  &   0.0000  &    0.0000  &   0.0042  &  4.9050\\
  $2$  &  -11.3972  &  -0.0008  &  -0.0002  &    0.0001  &   0.0014  &  6.2401\\
  $3$  &  -15.9939  &  -0.0190  &  -0.0001  &    0.0001  &   0.0289  &  9.7714\\
  $4$  &  -21.9017  &  -0.0438  &  -0.0002  &    0.0001  &   0.0603  &  14.0665 \\
  $5$  &  -28.3386  &  -0.0710  &  -0.0001  &    0.0000  &   0.0905  &  18.2002\\
  \hline
\end{tabular}}
\end{center}
\end{table}

%\begin{figure}[htp]
%  \begin{center}
%  \includegraphics[width=\textwidth]{Nu50.pdf}
%  \caption{Step input disturbance rejection with increasing $N_u$, disturbance occurs at $t=5.0s$.}
% \label{fi:Nu50}
%  \end{center}
%\end{figure}

\begin{figure}[htp]
  \begin{center}
  \includegraphics[width=\textwidth]{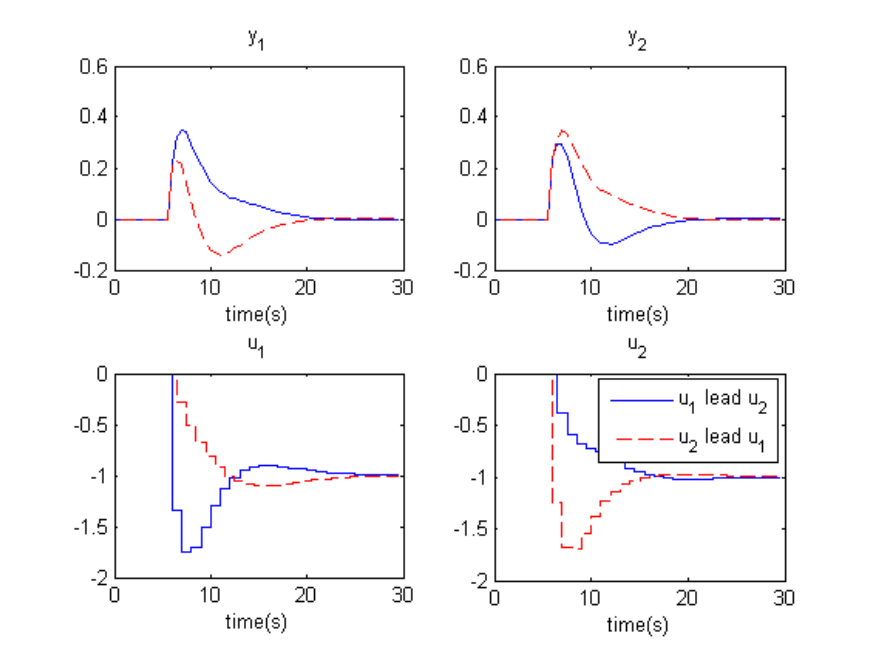}
  \caption{Effects of updating sequence, step input disturbance,
  $N_u=5$, solid: $(u_{1,k},u_{2,k},\ldots)$, dashed:$(u_{2,k},u_{1,k},\ldots)$.}
 \label{fi:DifSeqNu5}
 \end{center}
\end{figure}
%\bibitem{HoWK:07}
%Ho, W K, A Tay, M Chen, J Fu, HJ Lu and XC Shan, "Critical Dimension
%Uniformity via   Real-Time Photoresist Thickness Control". IEEE
%TRANSACTIONS ON SEMICONDUCTOR   MANUFACTURING,  20, no. 4 (2007):
%376-380.
%
%\bibitem{HoWK:08}
%Ho, W K, A Tay, J Fu, M CHEN and Y FENG, "Critical dimension and
%real-time temperature   control for warped wafers".  To appear in
%JOURNAL OF PROCESS CONTROL,  (2008)
%
%\bibitem{AwaSchKai:04}
%K. EI-Awady, C. Schapter, and T. Kailath "Programmable Thermal
%Processing Module for Semiconductor Substrates", {\it IEEE
%Transactions on Control System Technology}, vol. 12(4), 2004, pp
%493-509.
%\end{document}

%%% This is file "robustexamples.tex".
%%% Edited from ECC'07 paper ("ecc07_mmpc.tex") by JMM, 18.12.2007.
%%% ECC'07 paper was originally written mostly by AGR.

%\section{Results}
\label{sec:robustexamples}
\subsection{Robust MMPC: Spring-Mass Example}
This section considers the control of the simple mechanical system
shown in Fig.~\ref{fig:masses}. The system comprises four point
masses moving in one dimension. Each has mass of five units and is
connected to the adjacent masses by a spring of stiffness one unit.
\begin{figure}[htb]
\begin{center}
\includegraphics[width=0.3\textwidth]{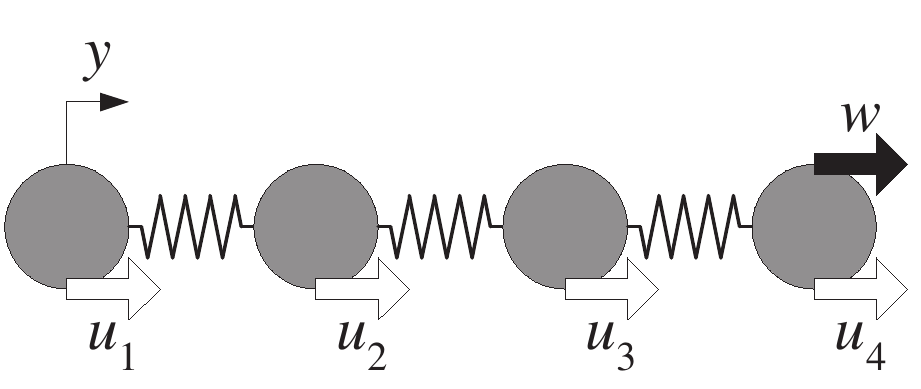}
\caption{Spring-Mass Example System} \label{fig:masses}
\end{center}
\end{figure}

Each controller minimizes control energy subject to a constraint on
the the position of mass~1, shown as output~$y$ in
Fig.~\ref{fig:masses}. Control energy is taken as $\int u(t)^T u(t)
dt$ over a 400s simulation. The inputs are the control moves~$\Delta
u_k$ applied to forces~$u_i$ acting on each mass, and therefore the
control force levels~$u(t)$ are elements in an augmented state
vector. All controllers were made robust to a disturbance force of
up to $0.01$~unit acting on mass~4. In the simulations, a
disturbance pulse was applied to that mass of magnitude~$0.01$ from
50s to 200s.

In the MMPC simulations, control moves were applied at intervals of one
second, \emph{i.e.}~channel~1 moved at $t=t_1$~seconds, then
channel~2 at $t=t_1+1$~seconds, and so on. In the comparison SMPC
simulations, moves were made on all channels every four seconds, but
to ensure fair comparison, the constraints were enforced at
intervals of one second as in MMPC. Computation time is taken as the
time spent in the ``quadprog'' function, totalled over all calls
during the simulation.

Figure~\ref{fig:massplots} shows the control input signals and the
output signals for each of the two controllers considered, using a
horizon of 120s in both cases. The asynchronous control moves can be
seen in the control signal plots from the MMPC simulation. In both
cases, the output signal runs tightly against the constraint (shown
dashed) for the duration of the disturbance pulse. This is as
expected, since the objective is to minimize control energy and
therefore the controller makes use of all available flexibility in
the output constraint. The output under MMPC is slightly further
from the limit than under SMPC, possibly because that controller
effectively solves a more constrained problem due to the reduced
decision variable set.  However, the effect is not significant.
\begin{figure}[t]
\begin{center}
%
%%\addtolength{\subfigcapskip}{-0.5\baselineskip}
%%\addtolength{\subfigtopskip}{-0.5\baselineskip}
%%\addtolength{\subfigbottomskip}{-0.5\baselineskip}
%
\def\picwidth{0.44\textwidth}
\subfigure[SMPC -
Controls]{\includegraphics[width=\picwidth]{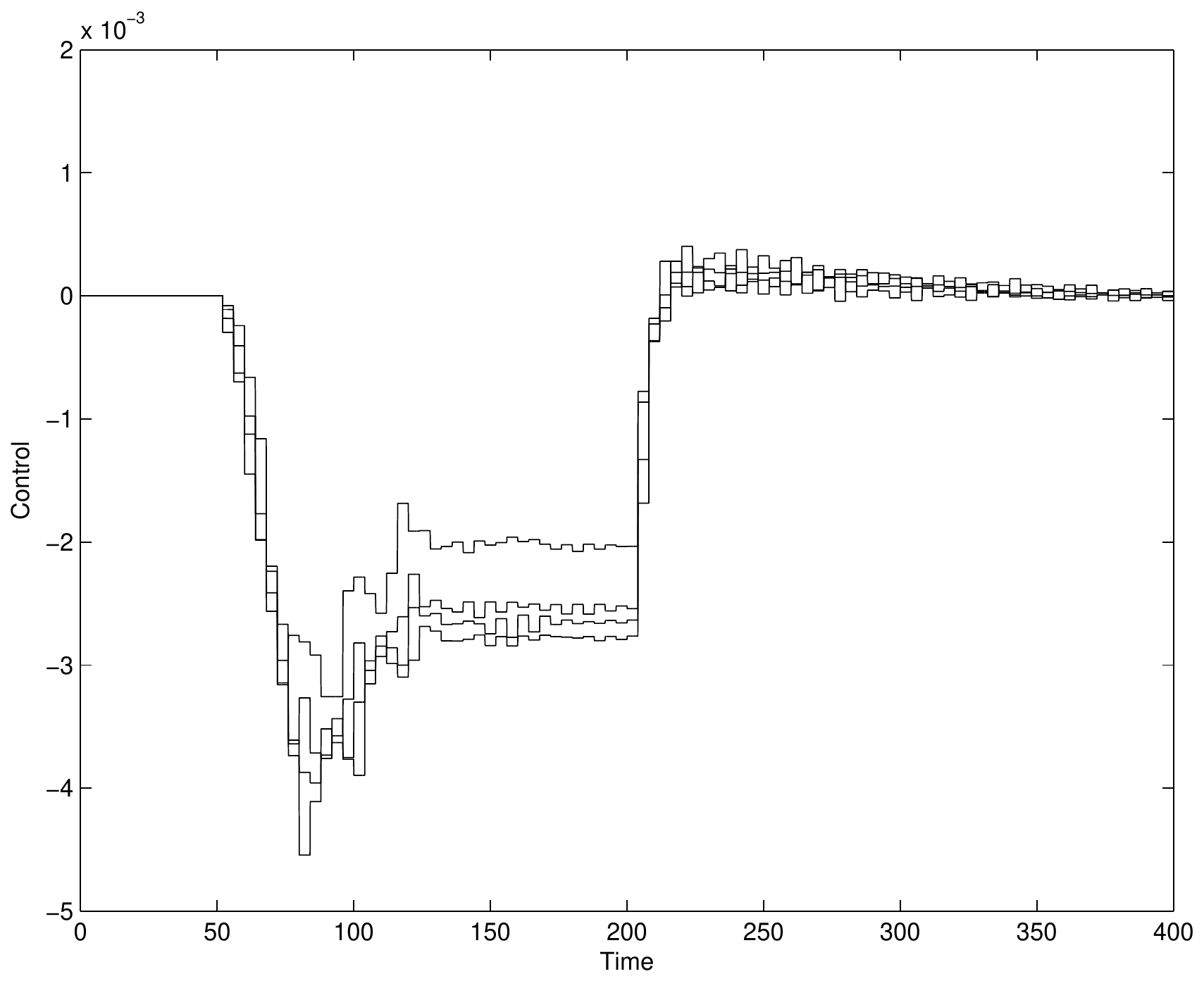}}
\subfigure[SMPC - Output]{\includegraphics[width=\picwidth]{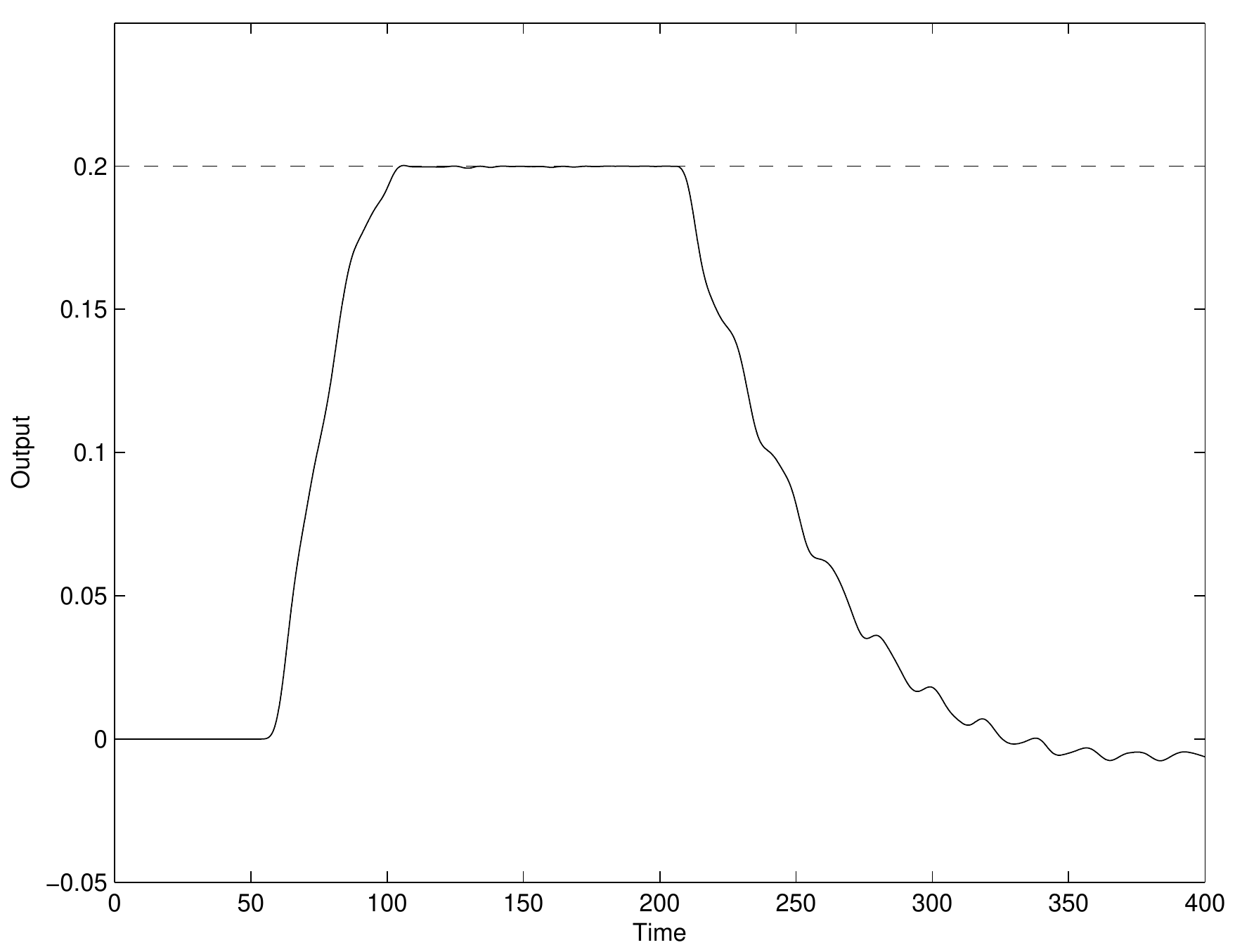}}\\
\subfigure[MMPC -
Controls]{\includegraphics[width=\picwidth]{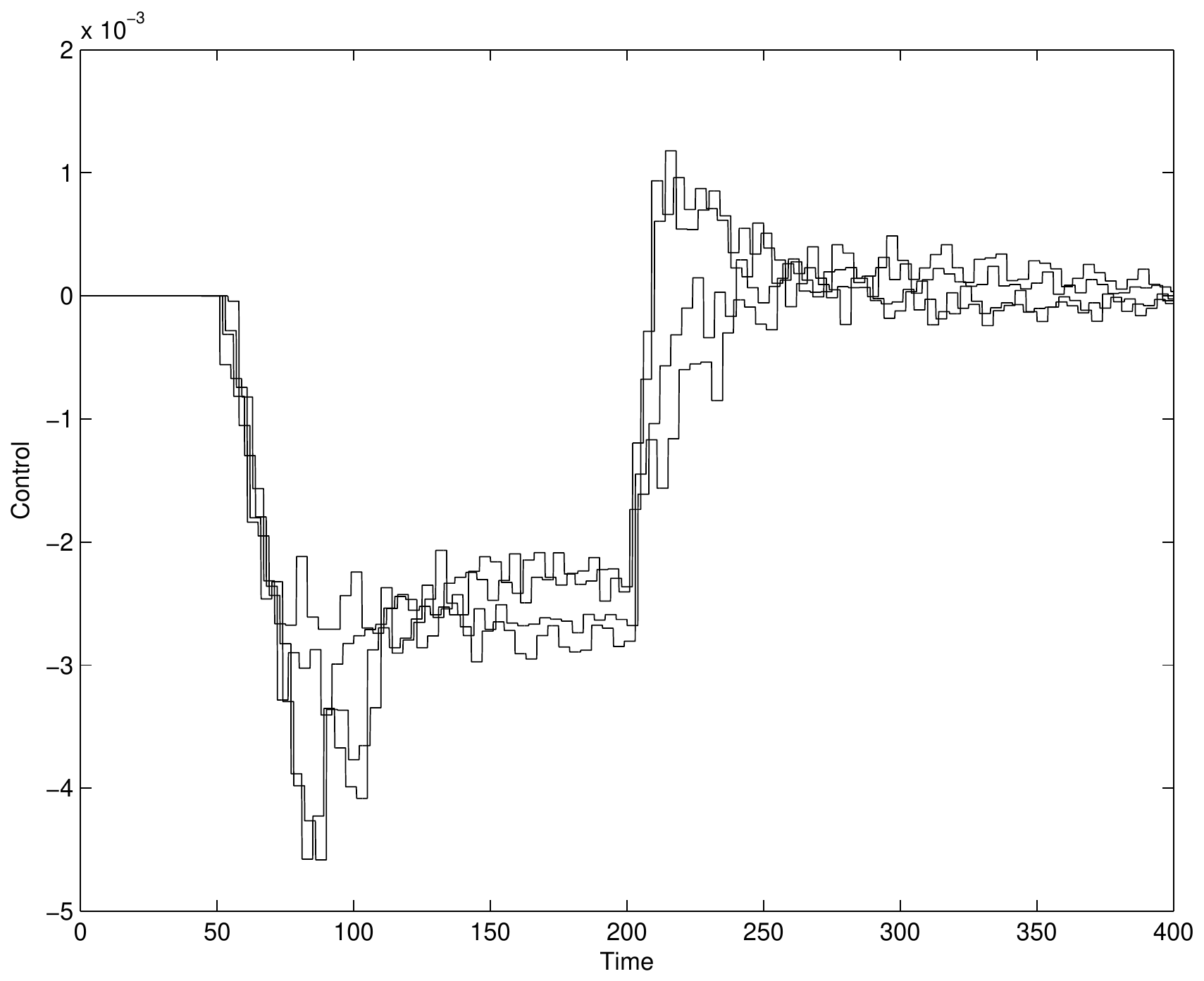}}
\subfigure[MMPC -
Output]{\includegraphics[width=\picwidth]{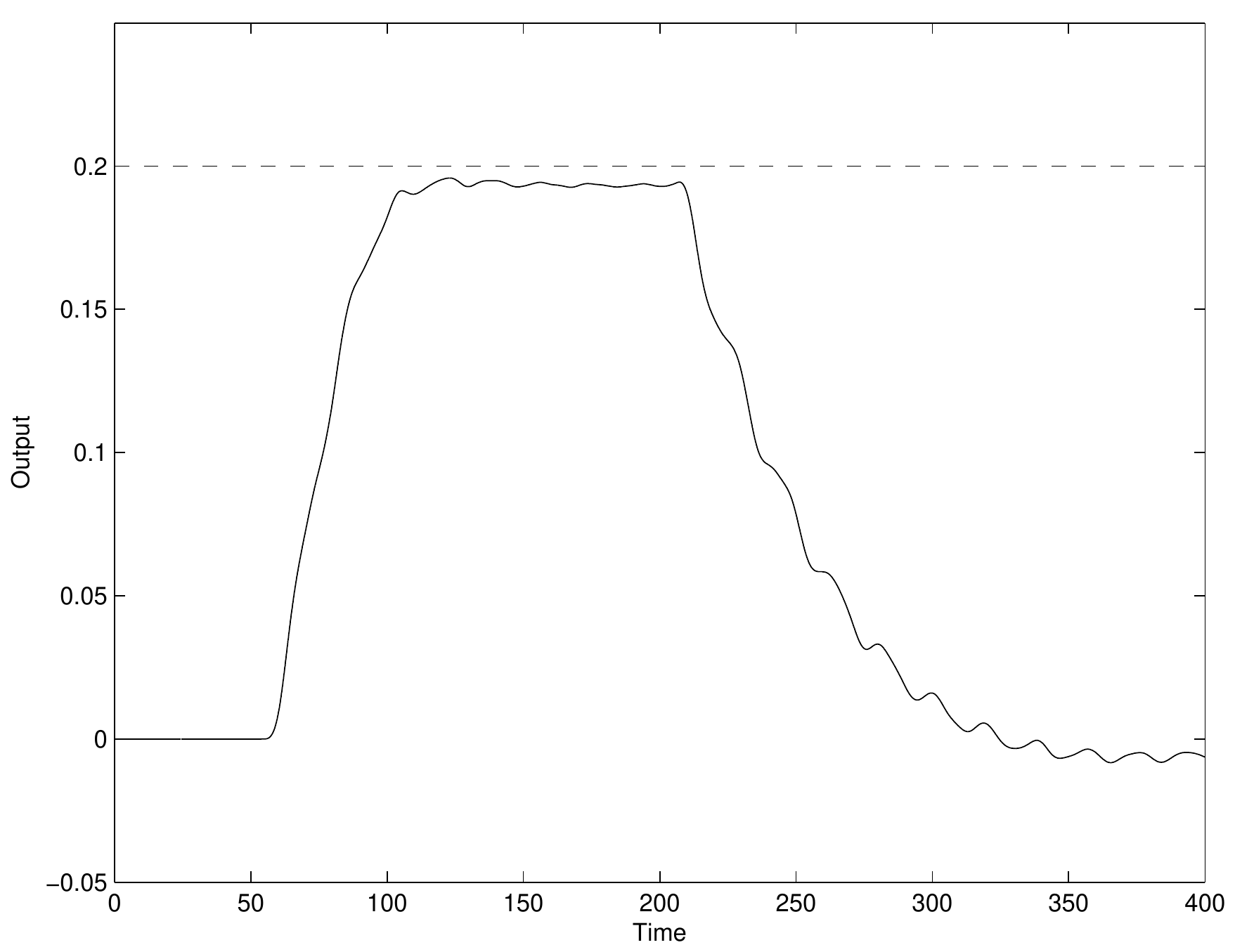}}
\caption{Spring-Mass Example: Responses to Disturbance Pulse}
\label{fig:massplots}
\end{center}
\end{figure}

To further illustrate the ability of the new robust MMPC to satisfy
hard constraints despite disturbances, the simulation using MMPC was
repeated using different constraint levels. The resulting output
signals are shown in Figure~\ref{fig:varcon}. In every case, the
signal goes right to its limit, but never beyond, and the
optimisations remain feasible. These results illustrate that the
constraints are active in these simulations and that the robust MMPC
method does not introduce undue conservatism.
\begin{figure}[t]
\begin{center}
\includegraphics[width=0.72\textwidth]{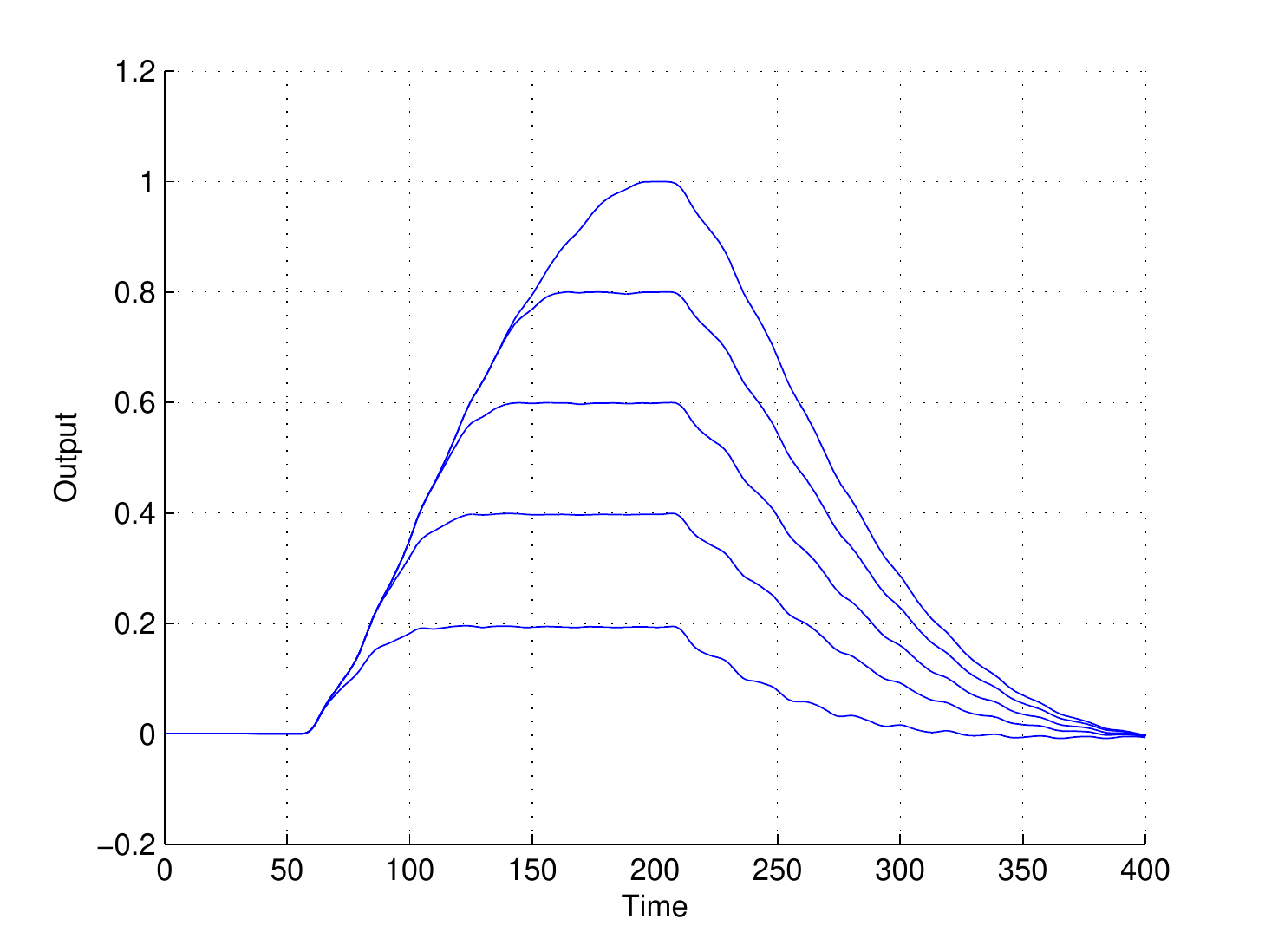}
\caption{Spring-Mass Example: Outputs from MMPC for
Constraint Settings 0.2, 0.4,
 0.6, 0.8, and~1.0}
\label{fig:varcon} \vspace{-0.1in}
\end{center}
\end{figure}

Table~\ref{tab:compmech} compares detailed statistics from the results
in Fig.~\ref{fig:massplots}. Observe that the performance, in terms of
the control energy, is roughly the same for both controllers. However,
MMPC is slightly faster than SMPC, since its sub-problems have only a
quarter as many decision variables as SMPC. This illustrates the
underlying premise of MMPC: it is faster to solve a sequence of four
problems of 31~variables than one problem of~124.
\begin{table}[t]
\begin{center}
\caption{Spring-Mass Example: Results for Each Controller Rejecting
Disturbance Pulse.} \label{tab:compmech}
\begin{tabular}{||l||c|c|c||}
\hline\hline
Controller & SMPC & MMPC \\
\hline\hline
$\int \mathbf{u}(t)^T \mathbf{u}(t) dt \times 1000$ & 4.312 & 4.320 \\
\hline
Computation Time (s) & 6.6 & 5.6 \\
\hline
N$^{\rm o.}$ of QP Solutions & 100 & 400 \\
\hline
N$^{\rm o.}$ of Decision Vars. per QP & 124 & 31 \\
\hline\hline
\end{tabular}
\end{center}
\end{table}

To further explore the issue of scalability, the simulations from
Fig.~\ref{fig:massplots} using SMPC and MMPC were repeated with
various horizon lengths. Figure~\ref{fig:varhoriz} shows the variation
of total computation time with horizon length for both
controllers. With a very short horizon, SMPC is faster than MMPC. We
hypothesize that this is due to overheads in the QP solver, such as
set-up time, which dominate the solution time for small problems and
therefore penalise the more frequent optimisation calls of MMPC.
However, as the horizon length increases, the computation time becomes
dominated by the actual solution process and MMPC scales more
favorably than SMPC.
\begin{figure}[t]
\begin{center}
\includegraphics[width=0.72\textwidth]{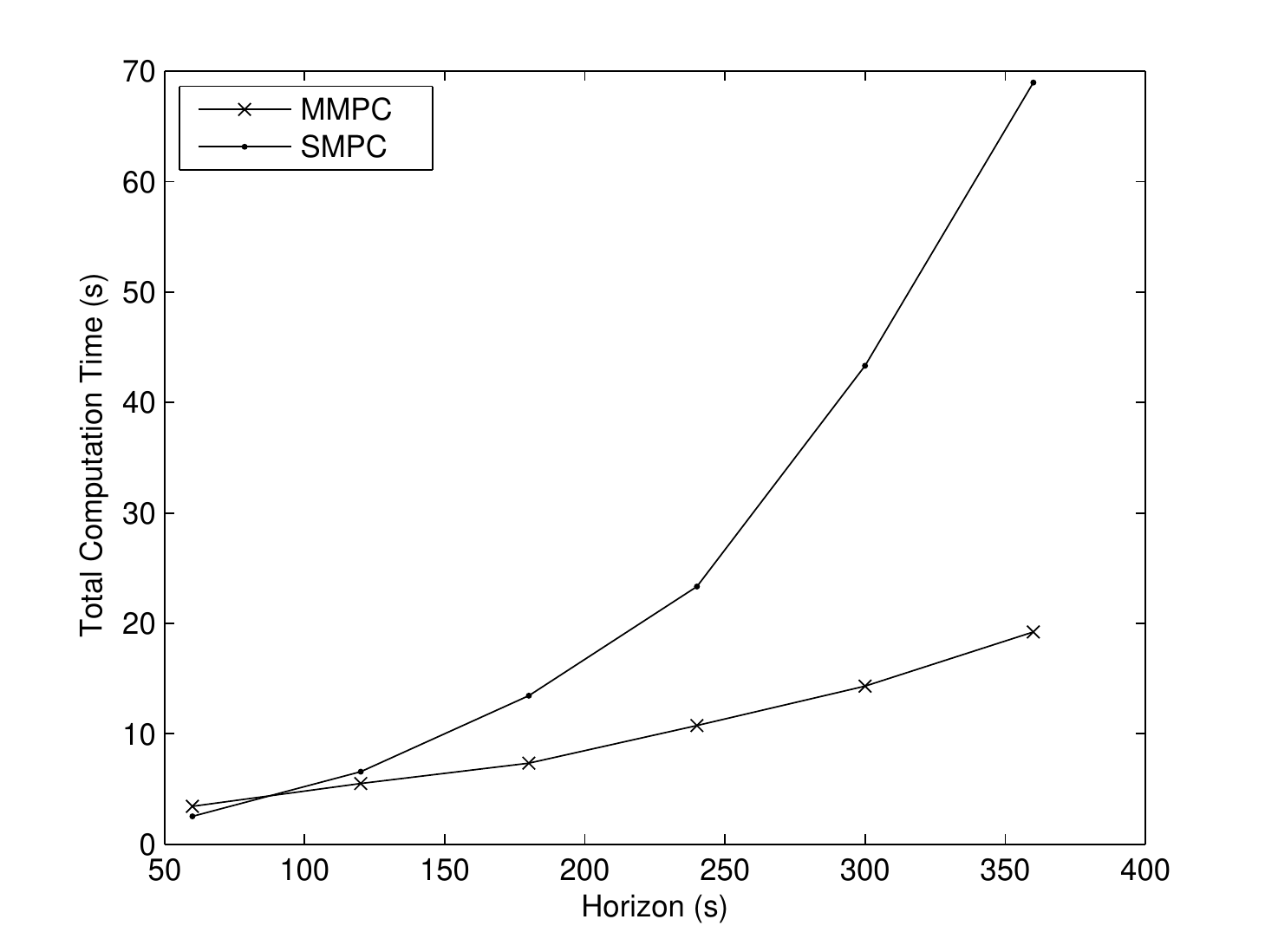}
\caption{Spring-Mass Example: Variation of Computation Time with
  Horizon Length for SMPC and MMPC}
\label{fig:varhoriz} \vspace{-0.1in}
\end{center}
\end{figure}

\subsection{Robust MMPC: Flight Dynamics Example}
This section considers longitudinal control of an A-7A Corsair~II
aircraft. The dynamics model was taken from Example~6.1 in
Ref.~\cite{cook} and augmented to include a thrust input as well as
the elevator input. Both inputs are constrained to~$[-0.04, 0.04]$
and the constraints are made robust to input disturbances in the
range~$[-0.01,0.01]$ on each channel. The simulation runs for 200s
and a disturbance of~$0.01$ is applied to both channels from~20s
to~120s. The planning horizon is 80s in all cases and the objective
is to minimize~$x_2^2$ where the state element $x_2$~corresponds to
the velocity normal to the aircraft axis in the body frame.

Figure~\ref{fig:acplots} shows the control and output signals from
simulations using the two different controllers. SMPC executes moves
on both channels at intervals of one second. MMPC performs a single
move on alternating channels every half a second.  Thus the total
number of moves on each channel in each simulation is the same.
Table~\ref{tab:compac} compares the results using the same metrics as
in the previous section, except for the performance which is here
taken as the peak value of the normal velocity~$\|x_2\|_\infty$.

Unlike in the spring-mass example, there is significant variation in
performance between the two controllers. The MMPC controller, with
its faster response time, is able to mitigate the short period
response more effectively than SMPC, which leaves a significant
spike at the onset of the disturbance, indicating that in this case,
it is better to respond to a disturbance quickly with one channel
than slowly with both. MMPC also requires significantly less
computational effort than SMPC for this example. Note that the
computation times are approximately in accordance with the expected
$O(\nu^3)$ behaviour, where $\nu$ is the number of decision
variables: in this example SMPC has 80 decision variables, and 200
QP problems are solved during the simulation, whereas MMPC has 41
variables, and 400 QP problems are solved. $(200\times
80^3):(400\times 41^3) = 3.7$, which is quite close to the ratio of
computation times $42.25:9.15 = 4.6$.
\begin{figure}[t]
\begin{center}
%
%%\addtolength{\subfigcapskip}{-0.5\baselineskip}
%%\addtolength{\subfigtopskip}{-0.5\baselineskip}
%%\addtolength{\subfigbottomskip}{-0.5\baselineskip}
%
\def\picwidth2{0.42\textwidth}
\subfigure[SMPC -
Controls]{\includegraphics[width=\picwidth2]{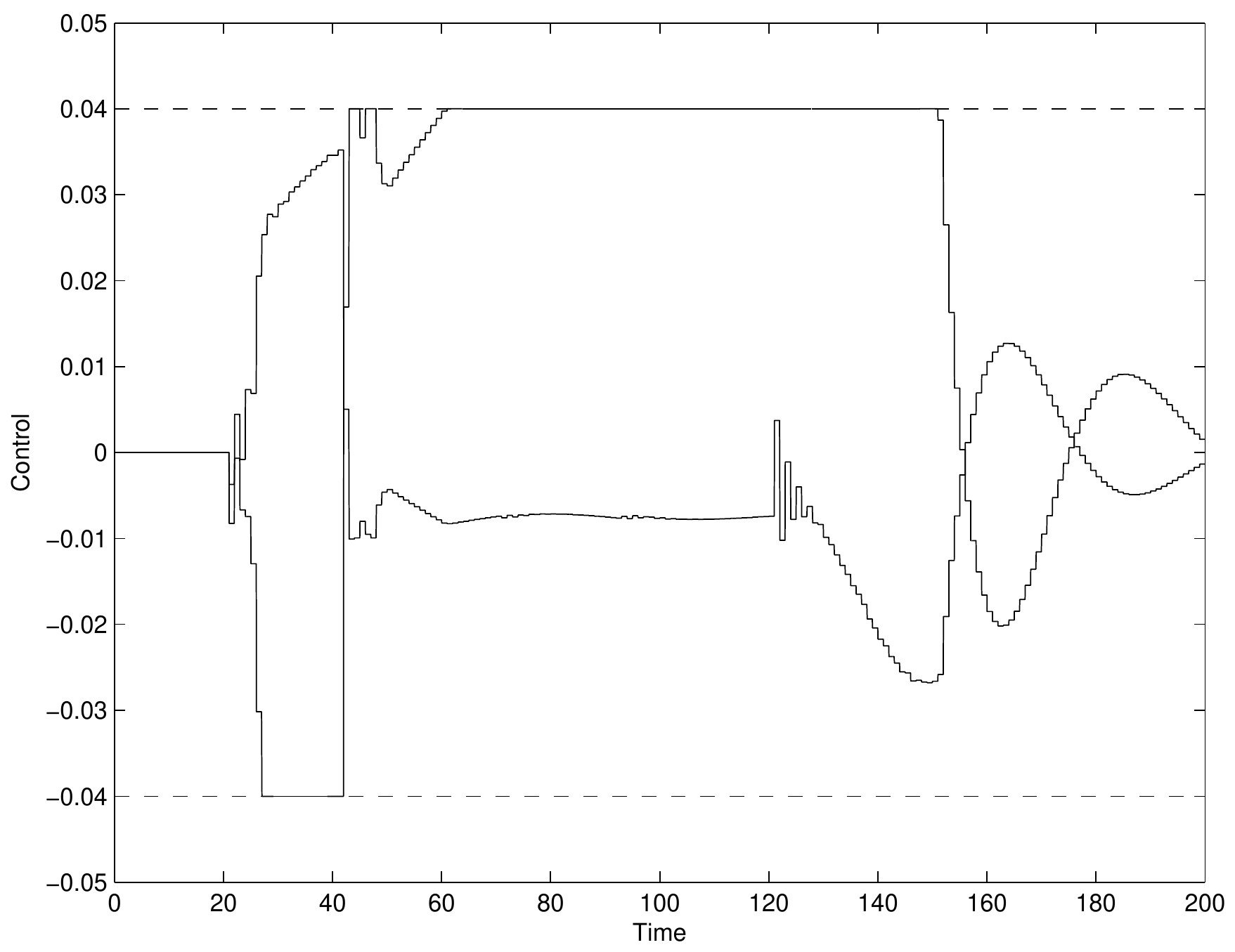}}
\subfigure[SMPC - Output]{\includegraphics[width=\picwidth2]{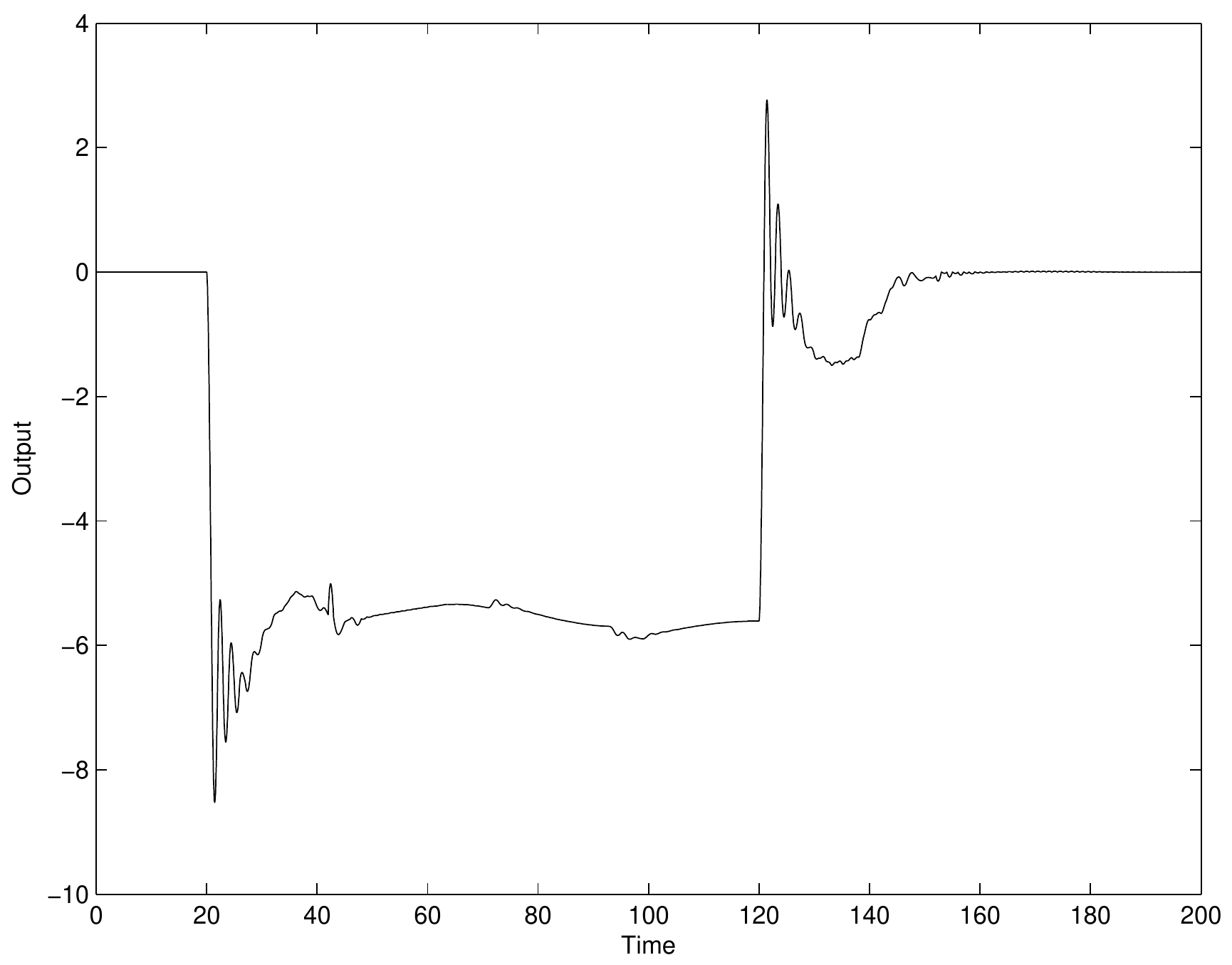}}\\
\subfigure[MMPC -
Controls]{\includegraphics[width=\picwidth2]{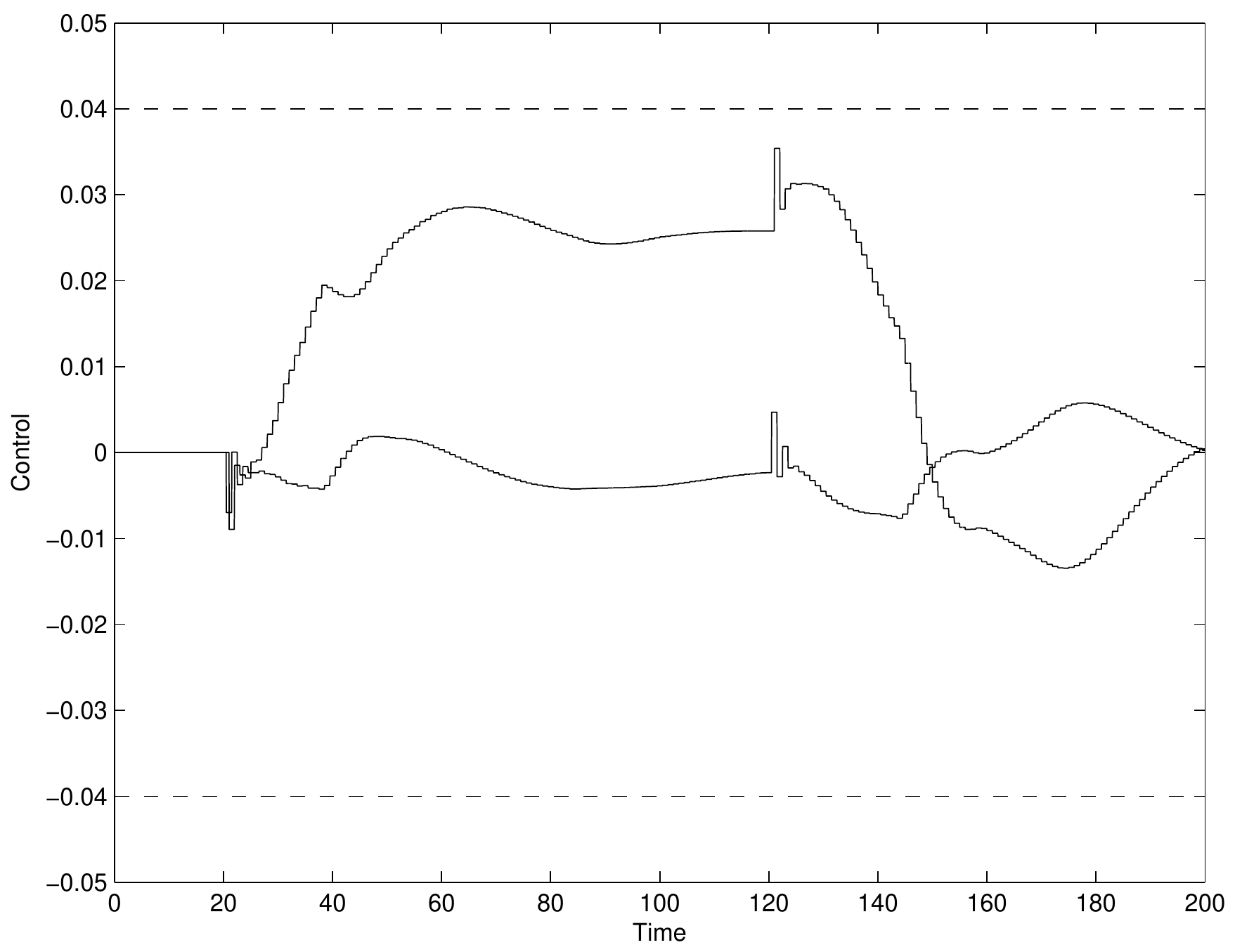}}
\subfigure[MMPC -
Output]{\includegraphics[width=\picwidth2]{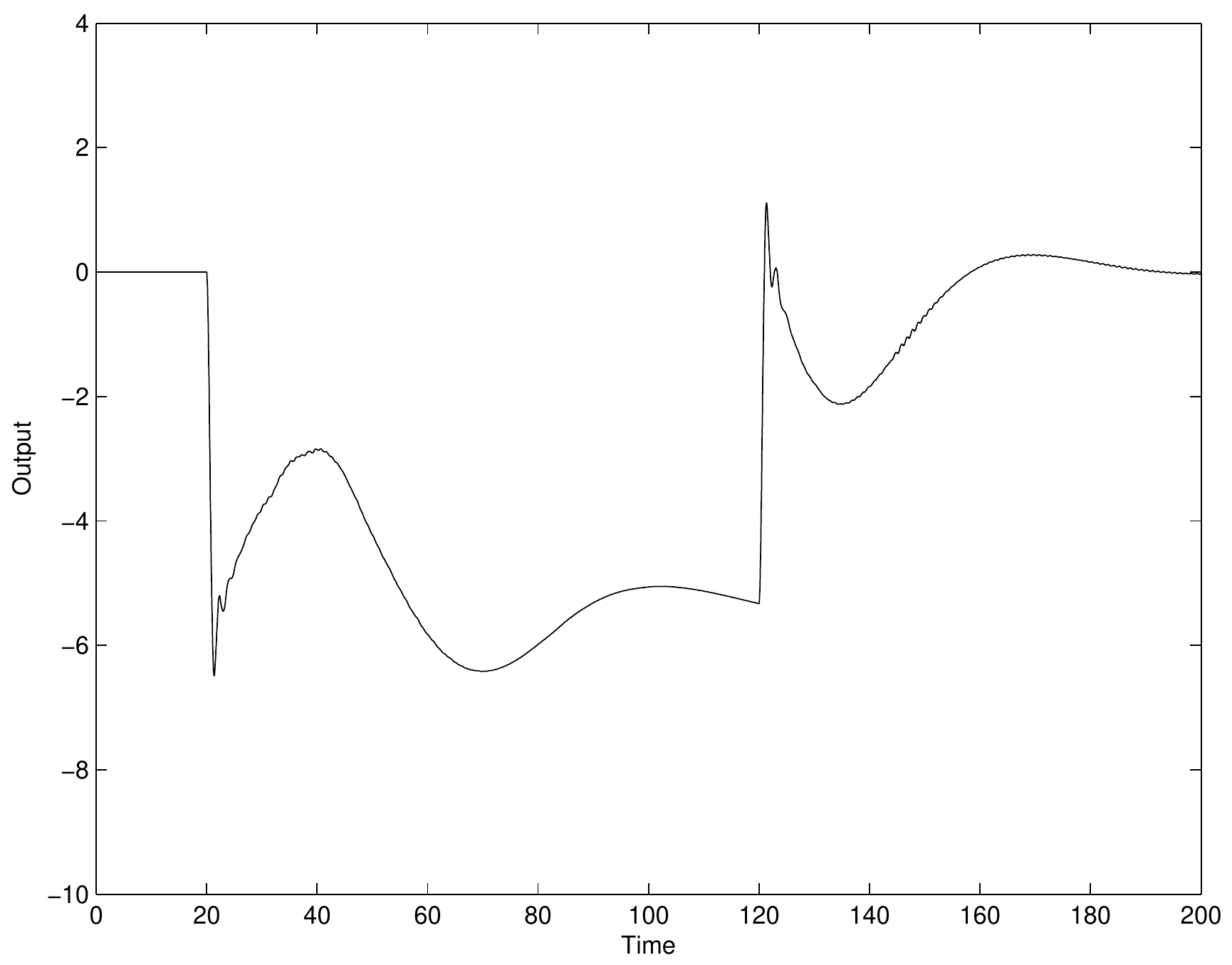}}
\caption{Aircraft Example: Responses to Disturbance Pulse}
\label{fig:acplots}
\end{center}
\end{figure}
\begin{table}[t]
\begin{center}
\caption{Aircraft Example: Results for Each Controller Rejecting
Disturbance Pulse.} \label{tab:compac}
\begin{tabular}{||l||c|c|c||}
\hline\hline
Controller & SMPC & MMPC \\
\hline\hline
$\|x_2(t)\|_\infty$ & 8.53 & 6.49 \\
\hline
Computation Time & 42.25 & 9.15 \\
\hline
N$^{\rm o.}$ of QP Solutions & 200 & 400 \\
\hline
N$^{\rm o.}$ of Decision Vars. per QP & 80 & 41 \\
\hline\hline
\end{tabular}
\end{center}
\end{table}

\section{Conclusion} \label{conclusion}

In this work %, two versions of
a novel control scheme known as \emph{Multiplexed} MPC was proposed,
which is %The second of these is
expected to be of practical benefit
because it offers reduced computational complexity.
%in those MPC applications for which
%computational complexity is a limiting factor.
%as an efficient method to
%cope with MPC applications where computational complexity may be
%an issue.
Multiplexed model predictive control (MMPC) updates one input at a
time, of a multi-input controlled plant. The motivation is to reduce
the computational complexity of MPC, in order to allow reduced
control update intervals. For some plants this leads to improved
control, as a result of the controller being able to react to
disturbances more quickly. MMPC scales well with increasing numbers
of inputs, since the computational complexity
% of the `Scheme 2' variant
depends only weakly on the number of inputs. The proposed MMPC
scheme has been proved to be nominally stable. The nominal stability
of a large class of other multiplexed MPC schemes follows by the
same argument as we used in this paper.

%It is interesting, and potentially important, to observe that the assumption of equal intervals
%between the updates of plant inputs is not essential to our proposal. Any pattern of update
%intervals can be supported, providing that it repeats in subsequent update cycles.

Some performance benefit over conventional MPC can be obtained as
a result of faster reactions to disturbances, despite suboptimal
solutions being obtained. This has been demonstrated by an
example.
%It has also been demonstrated that incorporating
%constraints in MMPCs can also improve performance.
However, the closed loop disturbance rejection performance under
MMPC is time varying because of the periodic nature of the control
scheme.
% A more detailed study on the effect difference input
%sequence has on the closed-loop performance is a topic of current
%research.

%the cost comparison between MMPC and a well designed
%SMPC\footnote{the SMPC uses the faster rate $x_{k+i}$ in the cost
%function} is indeterminate and depends on the initial conditions.

%A more detailed study on the selection of the sequencing of the
%inputs to In addition, the updating sequence of the control has an
%effect on the cost. A more detailed cost and performance
%comparison, especially in the presence of constraints and
%disturbances, between the MMPC and SMPC is a topic of current
%research.

%control scheme in disturbance rejection
%is illustrated through an example.

%\section{Conclusion} %%% The next 3 paras are the Conclusion of the ECC'07 paper.
%

In this paper we have extended the basic MMPC idea to obtain robust
feasibility and robust constraint satisfaction in the presence of
unknown but bounded disturbances.

Simulation examples have demonstrated that our scheme succeeds in
maintaining constraint satisfaction and feasibility despite the
presence of disturbances. Furthermore, they have shown that
performance improvements can indeed be obtained in some
circumstances, compared with conventional MPC, they have indicated
the kind of computational speed-up that can result from adoption of
the MMPC scheme, and they have illustrated that these benefits are
retained in circumstances where the constraints are active.

\bibliography{MPC,aeagr}
%\appendix

%\input{mmpcformula}
%\input{smpcformula}
%\input{MMPCcostAppendix}

%\input{mmpcaxb4Jul06}
%\input{costcomparison}
%\input{abstability}

\end{document}